\documentclass[bibyear]{aa} 

\usepackage{graphicx}
\usepackage{txfonts}
\usepackage{color}
\usepackage{sidecap}

\begin{document}

   \title{The \textit{Herschel} view of the nebula around the luminous\\ blue variable star AG Carinae
               \thanks{\textit{Herschel} is an ESA space observatory 
                with science instruments provided by European-led Principal 
                Investigator consortia and with important participation from
                NASA.}$^{,}$
                \thanks{Based in part on observations collected at the European
                Southern Observatory, La Silla, Chile}}

   \author{C. Vamvatira-Nakou\inst{1}
          \and D. Hutsem\'{e}kers\inst{1}\fnmsep\thanks{Senior Research 
               Associate FNRS}
          \and P. Royer\inst{2} 
          \and N.L.J. Cox\inst{2}
          \and Y. Naz\'e\inst{1}\fnmsep\thanks{Research Associate FNRS}
          \and G. Rauw\inst{1}
           \and \\C. Waelkens\inst{2} 
          \and M. A. T. Groenewegen\inst{3}
          }

   \institute{Institut d'Astrophysique et de G\'{e}ophysique, Universit\'{e}
              de Li\`ege, All\'{e}e du 
              6 ao\^ut, 17 - B\^at. B5c, B-4000 Li\`ege, Belgium \\
              \email{vamvatira@astro.ulg.ac.be}
           \and Instituut voor Sterrenkunde, KU Leuven, Celestijnenlaan 200D,
                   Bus 2401, B-3001 Leuven, Belgium
           \and Koninklijke Sterrenwacht van Belgi\"e, Ringlaan 3, B-1180 Brussels,
                 Belgium
             }

\date{Preprint online version: April 13, 2015}


\abstract{
Far-infrared {\it Herschel} PACS imaging and spectroscopic observations
of the nebula around the luminous blue variable (LBV) star AG Car have
been obtained along with optical imaging in the H$\alpha$+$[\ion{N}{ii}]$
filter.
In the infrared light, the nebula appears as a clumpy ring shell that
extends up to 1.2 pc with an inner radius of 0.4 pc. It coincides with
the H$\alpha$ nebula, but extends further out. Dust modeling of the
nebula was performed and indicates the presence of large grains. The
dust mass is estimated to be $\sim$ 0.2 M$_{\odot}$.
The infrared spectrum of the nebula consists of forbidden emission lines
over a dust continuum. Apart from ionized gas, these
lines also indicate the existence of neutral gas in a photodissociation
region that surrounds the ionized region. The abundance ratios point towards
enrichment by processed material. The total mass of the nebula ejected
from the central star amounts to $\sim$ 15 M$_{\odot}$, assuming a
dust-to-gas ratio typical of LBVs.
The abundances and the mass-loss rate were used to constrain the
evolutionary path of the central star and the epoch at which the nebula
was ejected, with the help of available evolutionary models. This
suggests an ejection during a cool LBV phase for a star of $\sim$ 55
M$_{\odot}$ with little rotation.
}

   \keywords{circumstellar matter --
             Stars: massive -- 
             Stars: mass-loss --
             Stars: variables: S Doradus --
             Stars: individual: AG Car}

   \authorrunning{C. Vamvatira-Nakou et al.}
   \titlerunning{The \textit{Herschel} view of the nebula around AG Car}

   \maketitle

%
\section{Introduction}
\label{sec:introduction}

The term ``luminous blue variable'' (LBV) was used by Conti (\cite{con84})
for the first time and referred to hot luminous massive variable stars
that are evolved, but are not Wolf-Rayet (WR) stars. Nowadays, LBVs, also
known as S Doradus variables, are considered to be massive evolved
stars mainly characterized by
a) high luminosity, $\sim10^{6}\ \mathrm{L}_{\odot}$; b) photometric
variability with amplitudes from $\sim$0.1 mag (small oscillations)
up to $\geq 2\ \mathrm{mag}$ (giant eruptions); and c) high mass-loss
rates $\sim10^{-5} - 10^ {-4}\ \mathrm{M}_{\odot}\ \mathrm{yr}^{-1}$
(Humphreys \& Davidson \cite{hum94}). Their location on the Hertzsprung-
Russell (HR) diagram is in the upper left part although some of them undergo
occasional excursions to the right.

An early-type O star with initial mass $\ge 30\ \mathrm{M}_{\odot}$ evolves
to a WR star by losing a significant fraction of its initial mass. Luminous blue variables
represent a short stage ($\sim10^{4} - 10^{5}\ \mathrm{yr}$) in this
evolutionary path according to current evolutionary scenarios (Maeder \&
Meynet \cite{maed10}). Although stellar winds can be responsible for
stellar mass-loss, the mass-loss rates of O stars have been revised downward
in the past few years (Bouret et al. \cite{bouret}; Fullerton et al.
\cite{fullerton}; Puls et al. \cite{puls}). Consequently, episodes of extreme
mass-loss during an intermediate evolutionary phase, like a LBV or a red
supergiant (RSG) phase, are now thought to play a key role, which is why
the study of the LBVs and their circumstellar environments is crucial for
understanding massive star evolution.

Such extreme mass-loss leads to the formation of ejected nebulae, which
have been observed around many LBVs (Hutsem\'{e}kers
\cite{hut94}; Nota et al. \cite{nota95}). They are classified into
three categories according to their morphology: shell nebulae, filamentary
nebulae and peculiar morphologies (Nota et al. \cite{nota95}). The study
of these circumstellar environments can reveal the mass-loss history of
the central star since they are formed by the material that has been
ejected from the central star in a previous evolutionary phase.
Dust and molecular gas (CO) have been revealed by infrared and millimeter
studies of LBV nebulae (McGregor et al. \cite{mcgr88}; Hutsem\'{e}kers
\cite{hut97}; Nota et al. \cite{nota02}).

Some LBVs are surrounded by more than one nebulae. This is the case of
the LBV G79.29+0.46. Near-infrared and millimeter data analyzed by
Jim\'{e}nez-Esteban et al. (\cite{jim10}) revealed multiple shells
around this star. Infrared observations by the $\textit{Herschel}$ Space
Observatory (Pilbratt et al. \cite{pilbratt}) revealed a second nebula
around the LBV WRAY 15-751 (Vamvatira-Nakou et al. \cite{vamv13}) 

AG Car (=HD 94910 =IRAS 10541-6011) is a well-studied prototypical LBV.
Its variability was first discovered by Wood (\cite{wood14}). It was
first classified as a P Cygni star by Cannon (\cite{can16}) and finally
classified as a LBV by Humphreys (\cite{hum89_1}). Numerous studies
show that this star exhibits photometric and spectroscopic variability.
In the optical V-band, the photometric changes during the S Dor
cycle are about 2 mag on a timescale of 5-10 years (Stahl
\cite{sta86_a}; van Genderen et al. \cite{van88}; Leitherer et al.
\cite{lei92}; Sterken et al. \cite{ster96}). In addition, smaller
variations of 0.1-0.5 mag on a timescale of about 1 year were discovered
(van Genderen et al. \cite{van97}). During the periods of visual minimum,
AG Car has a spectrum of a WR star, with Ofpe/WN9 spectral type
according to Stahl (\cite{sta86_a}) and with WN11 according to a more
recent study by Smith et al. (\cite{smith94}). During the periods of
visual maximum, AG Car's spectrum corresponds to that of an early-A
hypergiant (Wolf \& Stahl \cite{wol82}, Stahl et al. \cite{sta01}).

Humphreys et al. (\cite{hum89_2}) concluded that the distance to AG Car
is 6 $\pm$ 1 kpc, based on the calculated kinematic distance and on the
observed variation of the interstellar extinction with distance. This
result was confirmed by Hoekzema et al. (\cite{hoek92}), based again
on the extinction versus distance relation. Stahl et al. (\cite{sta01})
suggested a slightly lower distance of 5-6 kpc based on their calculations
of the heliocentric systemic velocity of AG Car (10 $\pm$ 5 km s$^{-1}$),
which is compatible with the value of Humphreys et al. (\cite{hum89_2})
considering the errors. Groh et al. (\cite{gro09}) calculated a similar
systemic velocity. Consequently, the value of 6 $\pm$ 1 kpc that
encompasses all measurements is adopted for all calculations in
this study.

Lamers et al. (\cite{lam89}) calculated $\log L/L_{\odot}$ = 6.2 $\pm$ 0.2
for the luminosity of AG Car and showed that it remains constant during
the light variations of the star, as was also found for other LBVs (R71: Wolf et al.
\cite{wol81}, R127: Stahl and Wolf \cite{sta86}). Later on, Leitherer et al.
(\cite{lei94}), in their study of the stellar wind of AG Car, found a slightly
lower bolometric luminosity of 6.0 $\pm$ 0.2 based on ultraviolet observations
combined with visual and near-infrared photometry. They also confirmed
the nonvariability of the bolometric luminosity during the S Dor cycle.
Given such a high value of bolometric luminosity, AG Car is well above
the Humphreys-Davidson limit (Humphreys and Davidson \cite{hum79}),
the limit above which a massive star becomes unstable and high mass
loss episodes take place.
However, in a recent study of the fundamental parameters of AG Car
during visual minima, Groh et al. (\cite{gro09}) concluded that the
bolometric luminosity of AG Car does change during the S Dor cycle. They
obtained a maximum value of the bolometric luminosity during minimum phase
of $\log L/L_{\odot}$ = 6.18, with a variation amplitude of
$\Delta(\log L/L_{\odot}) \sim$  0.17 dex. This luminosity variation lies
inside the limits of the previously calculated values considering the errors.
In all these studies, the distance 6 $\pm$ 1 kpc was used.

Thackeray (\cite{thak50}) discovered that AG Car is surrounded by a
nebulous shell that has the shape of an elliptical ring with nonuniform
intensity. He measured the size of the nebula to be 39$\arcsec\times$
30$\arcsec$ and the width of the ring to be about 5$\arcsec$.
Stahl (\cite{sta87}) studied direct CCD imaging data of AG Car and
suggested that very likely the nebula around this star is the
result of a heavy mass-loss episode that took place during an S Dor
outburst and not the result of interstellar material that was swept up
by the stellar wind.
Smith (\cite{smith91}) studied the dynamics of AG Car nebula based on
spectroscopic observations in the optical waveband. She measured an
average nebular expansion velocity of 70 km s$^{-1}$ and concluded that the
shell expansion is roughly symmetrical. She also reported the presence
of a jet-like bipolar mass outflow, expanding with a velocity of
83 km s$^{-1}$ and distorting the northeastern edge of the shell.
Nota et al. (\cite{nota92}) studied the nebula around AG Car using
high-resolution coronographic imaging and spectroscopic data so as
to constrain its geometry. They confirmed the value of the expansion
velocity of Smith (\cite{smith91}) and concluded that this nebular shell
shows a deviation from spherical symmetry, based on the observed radial
velocity variations and on the gas distribution observed in
the images.

Voors et al. (\cite{voo00}) studied AG Car in the infrared waveband by
modeling ground-based infrared images taken at about 10\ \mbox{$\mu$m}
and ISO spectroscopic observations, from which they derived the properties
of the circumstellar dust. The dust shell is detached
and slightly elongated. The ionized gas appears co-spatial with the dust.
Polycyclic aromatic hydrocarbons (PAHs) are present. The dust shell
contains mostly large grains, although very large grains are present
and also a population of small, warm, out of thermal equilibrium grains
that produce continuum and PAH bands emission.

Duncan and White (\cite{dun02}) observed the AG Car nebula at the radio
wavelengths (3 and 6 cm). Their 3 cm wavelength radio image revealed a
nebula with a detached ring shape, very similar to the morphology in the
H$\alpha$+$[\ion{N}{ii}]$ filter (see Sect.~\ref{sec:morphology of the
nebula}).

Nota et al. (\cite{nota02}) detected $ ^{12}CO\ J=1 \rightarrow0$ and
$J=2\rightarrow1$ emission from AG Car for the first time. The CO line
profiles indicate a region of molecular gas that is close to the star,
expanding slowly and not originating from the gaseous nebula. They
argued that the most plausible scenario to explain the observed profile
is the presence of a circumstellar disk.

Weis (\cite{weis08}), using deep H$\alpha$ imaging, reported the presence
of diffuse emission in the form of a cone-like extension to the north of
the AG Car nebula and concluded that it is clearly part of the nebula. It
extends up to 28$\arcsec$ outside the nebula, increasing its size by about
two times. It has a higher radial velocity than the ring.

In this study we present an analysis and discussion of the
images and the spectrum of the AG Car nebula taken by PACS (Photodetector
Array Camera and Spectrometer, Poglitsch et al. \cite{poglitsch}), which
is one of the three instruments on board the $\textit{Herschel}$ Space
Observatory. The paper is organized as
follows. In Sect.~\ref{sec:observations and data reduction} the observations
and the data reduction procedure are presented. A description of the nebular
morphology based on these observations is given in Sect.~\ref{sec:morphology
of the nebula}. The dust continuum emission is modeled in Sect.~\ref{sec:dust
continuum emission}, while the emission line spectrum is analyzed in
Sect.~\ref{sec:emission line spectrum}. In Sect.~\ref{sec:discussion} a
general discussion is presented and in Sect.~\ref{sec:conclusions}
conclusions are drawn.

\section{Observations and data reduction}
\label{sec:observations and data reduction}

\begin{figure*}[!]
\resizebox{\hsize}{!}{\includegraphics*{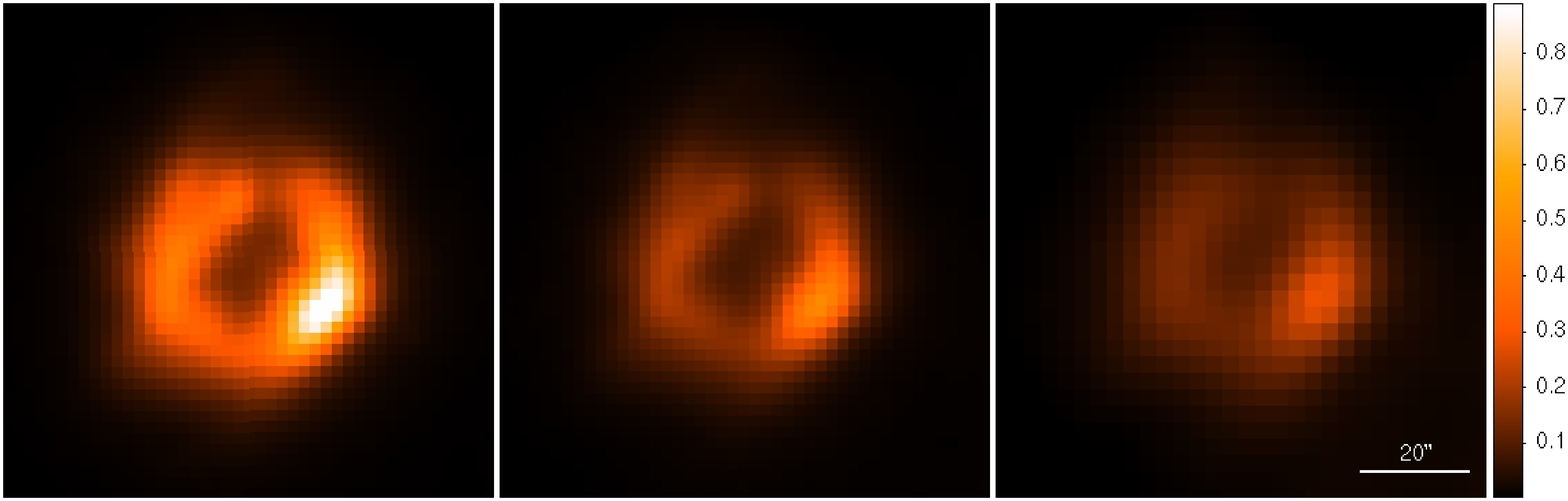}}\\%
\resizebox{\hsize}{!}{\includegraphics*{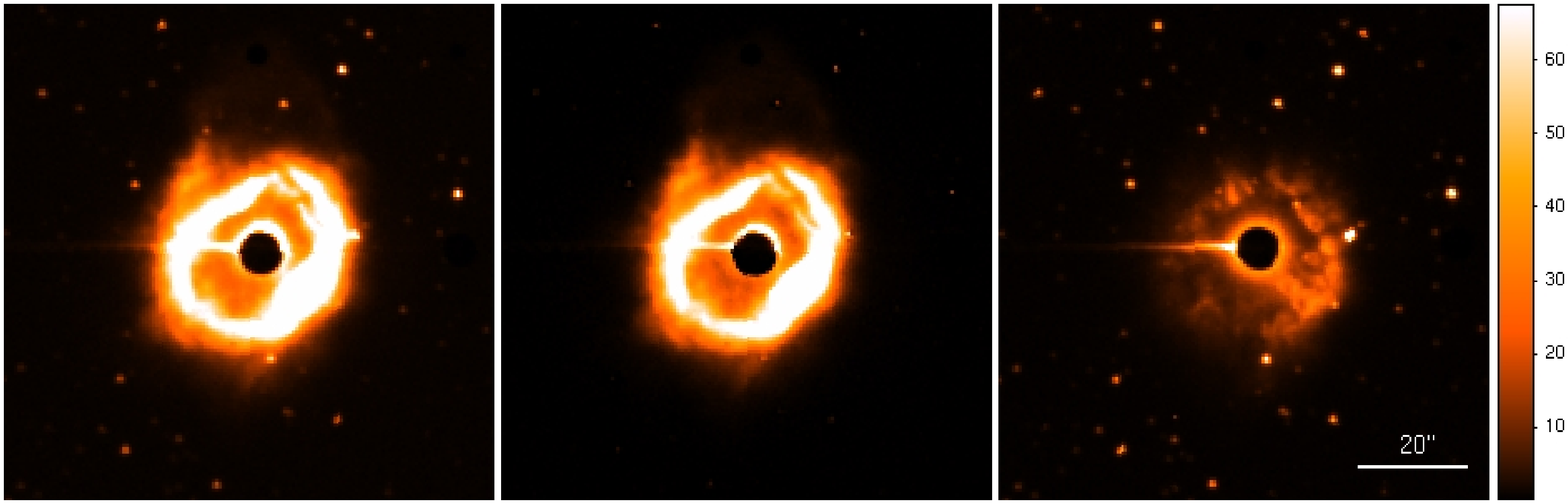}}
\caption{Images of the nebula around AG Car. Top: PACS images at 70\
\mbox{$\mu$m}, 100\ \mbox{$\mu$m,} and 160\ \mbox{$\mu$m} from left to
right. Bottom: the H$\alpha$+$[\ion{N}{ii}]$ image (left), the continuum
image (right) and the image resulting from the  subtraction of the continuum
image from the  H$\alpha$+$[\ion{N}{ii}]$ image after correcting for the
position offsets and for the different filter transmissions using field
stars (middle). The size of each image is 1.5$\arcmin\times1.5\arcmin$.
The scale on the right corresponds to the surface brightness (arbitrary
units). North is up and east is to the left.}
\label{imag}
\end{figure*}

\subsection{Infrared observations}

The infrared observations include imaging and spectroscopy of the AG Car
nebula and were carried out using PACS in the framework of the \textit{Mass-loss
of Evolved StarS (MESS)} Guaranteed Time Key Program (Groenewegen et al.
\cite{groenewegen}).

\begin{figure*}
\sidecaption
  \includegraphics[height=5.75cm]{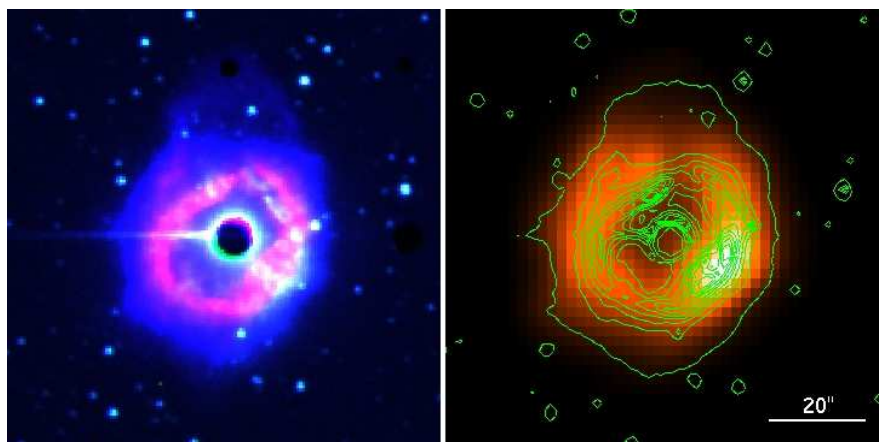}
     \caption{Left: View of the nebula in the optical. The bright
H$\alpha$+$[\ion{N}{ii}]$ ring nebula is illustrated in red (also shown in
Fig.~\ref{imag} at the same scale) while the fainter H$\alpha$+$[\ion{N}{ii}]$
emission is shown in blue, differently scaled in surface brightness, revealing
the northern extension (Weis \cite{weis08}). The continuum image that outlines
the dust scattered nebula is presented in green. Right: Contour image
of the optical emission from the nebula (green lines) superposed on the infrared
image of the nebula at 70\ \mbox{$\mu$m} (also shown in Fig.~\ref{imag} at the
same scale). The size of each image is 1.5$\arcmin\times1.5\arcmin$.
North is up and east is to the left.}
     \label{agcar_faint}
\end{figure*}

The PACS imaging observations were carried out on August 12, 2010, which
corresponds to $\textit{Herschel}$'s observational day (OD) 456. The observing mode
was the scan map in which the telescope slews at constant speed (in our
case the ``medium'' speed of $20\arcsec/\mbox{s}$) along parallel lines in order
to cover the required area of the sky. Two orthogonal scan maps were obtained for
each filter. Our final data set consists of maps at 70, 100 and 160\ \mbox{$\mu$m}.
The observation identification numbers (obsID) of the four scans are
1342202927, 1342202928, 1342202929, and 1342202930; each scan has a duration of
157s.

To perform the data reduction we used the Herschel Interactive
Processing Environment (HIPE, Ott \cite{ott}) up to level 1. Subsequently,
the data were reduced and combined using the Scanamorphos software
(Roussel \cite{rous12}). The pixel size in the final maps
is 2$\arcsec$ in the blue channel (70, 100 $\mu$m) and 3$\arcsec$ in
the red channel (160 $\mu$m). It should be mentioned that the
$\textit{Herschel}$ PACS point spread function (PSF) full widths at
half maximum (FWHMs) are 5$\farcs$2, 7$\farcs$7, and 12$\arcsec$
at 70\ \mbox{$\mu$m}, 100\ \mbox{$\mu$m} and 160\ \mbox{$\mu$m},
respectively.

The spectrum of the AG Car nebula was taken on June 5, 2010
(OD 387), with the PACS integral-field spectrometer that covers the wavelength
range from 52\ \mbox{$\mu$m} to 220\ \mbox{$\mu$m} in two channels that
operate simultaneously in the blue 52-98\ \mbox{$\mu$m} band (second
order: B2A 52-73\ \mbox{$\mu$m} and B2B 70-105\ \mbox{$\mu$m}), and the
red 102-220\ \mbox{$\mu$m} band (first order: R1A 133-220\ \mbox{$\mu$m}
and R1B 102-203\ \mbox{$\mu$m}). Simultaneous imaging of a $47\arcsec \times
47\arcsec$ field of view is provided that is resolved in $5\times5$
square spatial pixels (i.e., spaxels). The two-dimensional field of view
is then rearranged along a $1\times25$ pixel entrance slit for the
grating via an image slicer employing reflective optics. Its resolving
power is $\lambda/\delta\lambda \sim 940-5500$ depending on the wavelength.
The observing template was the spectral energy distribution (SED) that 
provides a complete coverage between 52 and 220\ \mbox{$\mu$m}.
The two obsIDs are 1342197792 and 1342197793. HIPE was also used for the
data reduction. The standard reduction steps were followed and
in particular the subtraction of the background spectrum obtained through
chopping and nodding.

\begin{figure}[!]
\resizebox{\hsize}{!}{\includegraphics*{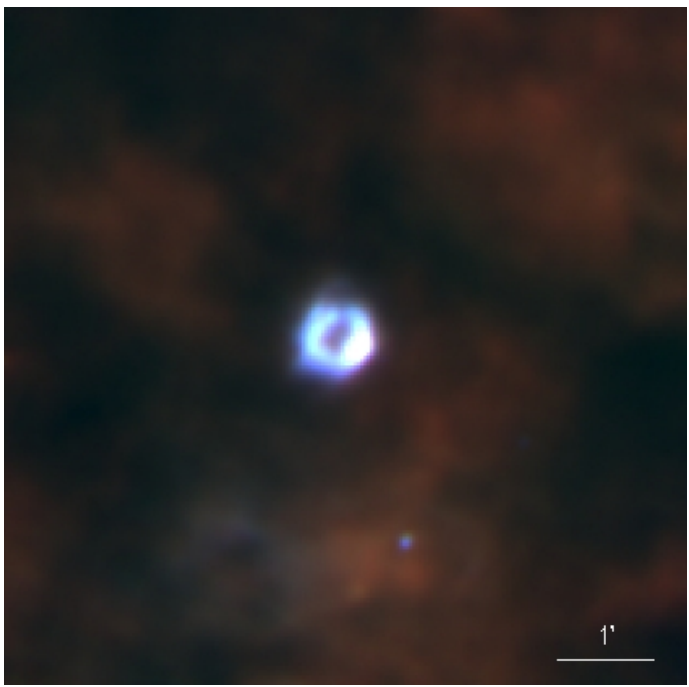}}
\caption{Three-color image (70 \ \mbox{$\mu$m} in blue, 100 \ \mbox{$\mu$m}
in green and 160 \ \mbox{$\mu$m} in red) of the AG Car nebula.
The LBV nebula appears located inside a cavity in the interstellar
medium. The size of the image is 7$\arcmin\times7\arcmin$. North is
up and east is to the left.}
\label{agcar_RGB}
\end{figure}

\subsection{Visible observations}

The optical images of the AG Car nebulae were obtained on April 6,
1995, with the 3.6 m telescope at the European Southern Observatory
(ESO), La Silla, Chile. A series of short (1s-10s) and longer
(30s-60s) exposures were secured in a H$\alpha$+$[\ion{N}{ii}]$ filter
($\lambda_{\rm c}$ = 6560.5\AA ; {\sc FWHM} =62.2\AA ) and in a continuum
filter just redwards ($\lambda_{\rm c}$ = 6644.7\AA; {\sc FWHM} =61.0\AA ).
The EFOSC1 camera was used in its coronographic mode: for the longer exposures,
the 8$\arcsec$ circular coronographic mask was inserted in the aperture wheel
and positioned on the central star while the Lyot stop was inserted in
the grism wheel (Melnick et al. \cite{mel89}). The frames were bias-corrected
and flat-fielded. The CCD pixel size was 0$\farcs$605 on the sky. The
night was photometric and the seeing around 1$\farcs$2. In order to
properly calibrate the images, three spectrophotometric standard stars
and three planetary nebulae with known H$\alpha$ flux were observed in the
H$\alpha$+$[\ion{N}{ii}]$ filter.

\section{Morphology of the nebula}
\label{sec:morphology of the nebula}

The three PACS infrared images of the nebula around the LBV AG Car at
70\ \mbox{$\mu$m}, 100\ \mbox{$\mu$m}, and 160\ \mbox{$\mu$m} along
with images taken at optical wavelengths are shown in Figs.~\ref{imag}
and ~\ref{agcar_faint}.

The infrared images reveal a dusty shell nebula with a clumpy ring
morphology that is clearly detached from the central star (not visible
at these wavelengths).  The brightness of the nebula is not uniform,
with the southwestern part being the brightest one. This agrees with
the nonuniform brightness of the images at 12.5\ \mbox{$\mu$m} and
12.78\ \mbox{$\mu$m} analyzed by Voors et al. (\cite{voo00}). An axis
at position angle (PA) $\sim$ 160\degr\ can be defined (north:
PA = 0\degr; east: PA = 90\degr). Cuts of the
70 $\mu$m PACS image through PA $\sim$ 160\degr\ and PA $\sim$ 70\degr\
(Fig.~\ref{synt_cuts}) show peaks at a radius of about 14$\arcsec$,
although with different intensities.  The ring extends up to
$\sim42\arcsec$. These values correspond to 0.4 pc and 1.2 pc,
respectively, at a distance of 6 kpc. On top of this central circular
nebula, there is a northern faint extension. No other more extended
nebula associated with the star can be detected. A three-color infrared
image of the nebula and its environment is illustrated in
Fig.~\ref{agcar_RGB}, which shows best this faint northern extension first
reported by Weis (\cite{weis08}), although the resolution is not good
enough to reveal its detailed structure. The LBV nebula seems to
be located in a cavity in the interstellar medium, probably excavated
by the star in a previous evolutionary phase. The radius of this empty
cavity is about 2.5$\arcmin$, which corresponds to 4.4 pc at a
distance of 6 kpc.

The H$\alpha$+$[\ion{N}{ii}]$ image (Fig.~\ref{imag}) shows that the
gas nebula around AG Car forms an elliptical shell with an average
outer radius of $\sim20\arcsec$ and an inner radius of
$\sim$11$\arcsec$. To more accurately investigate the morphology of the nebula
in the optical, a three-color image is presented in
Fig.~\ref{agcar_faint}.  The fainter nebular optical emission reveals
the northern extension described in Weis (\cite{weis08}). The 
nebula that surrounds the bright ring extends up to $\sim23\arcsec$,
while the north extension goes up to $\sim36\arcsec$ from the center
of the nebula. These numbers correspond to 0.7 pc and 1 pc,
respectively, at a distance of 6 kpc, in agreement with the size of
the nebula given in Weis (\cite{weis08}, \cite{weis11}).

In the H$\alpha$ light, the dynamics points to a spherically expanding
shell distorted by a more extended bipolar nebula (Smith \cite{smith91}; Nota
et al.  \cite{nota92}). In projection on the sky, the shell appears as
an elliptical ring with PA $\sim$ 131\degr, different from the
infrared shell PA.  Nevertheless, the contour image of the optical
emission (both bright and faint) superimposed on the infrared
image of the nebula at 70\ \mbox{$\mu$m} as illustrated in the right
panel of Fig.~\ref{agcar_faint} shows that the overall morphology of
the gas nebula is similar to the infrared dust morphology, although the
H$\alpha$+$[\ion{N}{ii}]$ ring nebula appears slightly smaller and
more elliptical than the infrared nebula.  The bright region at the
southwestern part of the gas nebula coincides with the bright region
of the infrared dust nebula. The northern faint extended structure,
which appears both in the infrared and in the optical, is likely a
lobe of the bipolar nebula. The extension seen to the south in the
H$\alpha$+$[\ion{N}{ii}]$ map and in the velocity maps of Smith
(\cite{smith91}) could constitute a part of a second fainter
lobe. The system may thus consist of a typical bipolar nebula seen
roughly through the poles (i.e., at inclination $\lesssim$ 30\degr),
with two faint lobes and a bright waist.

The nebula is also clearly detected in the optical continuum filter
(Fig.~\ref{imag}), indicating significant dust scattering. The
morphology of the dust reflection nebula appears somewhat different
from the morphology of the H$\alpha$+$[\ion{N}{ii}]$ emission.  This
reflection nebula around AG Car was first described in Paresce and
Nota (\cite{par89}) and its stunning structure resolved with the
Hubble Space Telescope (Nota et al. \cite{nota95}, \cite{nota96}).
More specifically, in the optical continuum image (Figs.~\ref{imag}
and ~\ref{agcar_faint}) the ring appears circular and very clumpy with
a jet-like feature that starts from the central part of the nebula
and extends towards the southwestern part, which is the brightest region
of the dust and gas emission. The northern extension of the nebula is
not detected in the optical continuum. As does the
H$\alpha$+$[\ion{N}{ii}]$ ring nebula, the optical continuum emission
appears slightly inside the infrared continuum emission, although a
detailed comparison is difficult given the lower spatial resolution of
the PACS images. The difference in morphology between the optical
continuum and the H$\alpha$+$[\ion{N}{ii}]$ nebulae may arise from
anisotropic illumination and thus ionization of different parts of
the nebula owing to its clumpy structure.

Considering the nebular expansion velocity, ${\rm v}_{\mathrm{exp}}$,
of 70 km s$^{-1}$ measured by Smith (\cite{smith91}), the kinematic age,
$t_{\mathrm{kin}}$, of the nebula can be estimated. As mentioned
above, the nebula in the infrared extends to 1.2 pc in radius,
\textit{r}, so it has a kinematic age of $t_{\mathrm{kin}} = r /{\rm
v}_{\mathrm{exp}}$ = 1.7$\times10^4$ years. The temporal difference
between the inner and the outer radius of the nebula is
1.1$\times10^4$ years.

\section{Dust continuum emission}
\label{sec:dust continuum emission}

Integrated flux densities were derived for the nebular shell at the three
PACS wavelengths by performing aperture photometry on the PACS images.
We also used imaging data taken from the archives of the Infrared Astronomical
Satellite (IRAS) mission (Neugebauer et al. \cite{neug84}) and the Infrared
Astronomical Mission AKARI (Murakami et al. \cite{mura07}). In these archival
data we did not include the IRAS observation at 12 $\mu$m because it is
only an upper limit or at 100 $\mu$m because it is not of high quality.
We note that the beam size of the IRAS and AKARI observations is large enough
to fully encompass the ring nebula around AG Car. Images from the SPIRE
(Spectral and Photometric Imaging Receiver, Griffin et al. \cite{grif10})
instrument on board the $\textit{Herschel}$ Space Observatory were also
included. They were taken from the observations of the $\textit{Herschel}$
Infrared Galactic Plane survey (Hi-GAL, Molinari et al. \cite{mol10}) made
immediately public for legacy. The Hi-GAL observations of the field around
AG Car were retrieved from the archive processed up to level 2. Only the maps
at 250 $\mu$m and at 350 $\mu$m were used, because at 500 $\mu$m the nebula
is very faint so that flux determination is highly uncertain. Integrated flux
densities were derived for the nebular shell at these two SPIRE wavelengths
by performing aperture photometry on the images.

We applied photometric color correction to all flux densities derived from
the observations of these three space missions. With this correction,
the monochromatic flux densities that refer to a constant energy spectrum
are converted to the true object SED flux densities at the photometric
reference wavelengths of each instrument.
We used the flux density ratio to derive the color temperature for
the color correction of the IRAS data and then we chose the corresponding
color-correction factor (Beichman et al. \cite{beich88}). The ratio
R (25,60) corresponds to a temperature of 140 K, so the correction
factors for this temperature were used. To color correct the AKARI FIS
and IRC data we fitted a blackbody to the two data sets independently
using the 25 $\mu$m IRAS observation because we needed a measurement
near the maximum of the curve. These fits led us to adopt the
color-correction factors that correspond to a temperature of 150
K for FIS (Yamamura et al. \cite{yam10}) and 220 K for IRC data
(Lorente et al. \cite{lorente08}). To estimate the color correction of
the Herschel-PACS data, we fitted a blackbody, considering again
the 25 $\mu$m IRAS observation. This fit gave a temperature of 150 K,
for which we adopted the corresponding correction factor (M\"{u}ller
et al. \cite{muller}). For the SPIRE data, the instructions for color correction
given in the SPIRE Handbook \footnote{http://herschel.esac.esa.int/Docs
/SPIRE/spire\textunderscore handbook.pdf}, were followed.

\begin{table}[t]
\caption{Color-corrected nebular flux densities.}
\label{table:1}
\centering
\begin{tabular}{l c c c c c}
\hline\hline                            \\
Spacecraft-Instrument &    Date & $\lambda$  & $F_{\nu}$ & Error  \\
                      &         & ($\mu$m)   & (Jy)      & (Jy)   \\
\hline\hline
IRAS\tablefootmark{a} &   1983  &   25       & 187.5  & 9.4   \\
                      &         &   60       & 177.7  & 28.4  \\
\hline
AKARI-IRC
\tablefootmark{b}     &  2007   &   9        & 9.04   & 0.14  \\
                      &         &   18       & 119.2  & 0.24  \\ 
AKARI-FIS
\tablefootmark{c}     &  2007   &   65       & 229.3  & 6.53  \\
                      &         &   90       & 81.0   & 3.9   \\
                      &         &   140      & 45.5   & 5.2   \\
                      &         &   160      & 35.1   & 4.6   \\
\hline
Herschel-PACS
\tablefootmark{d}     &  2010   &   70       & 173    & 2   \\
                      &         &   100      & 103    & 3   \\
                      &         &   160      & 42     & 3   \\
Herschel-SPIRE
\tablefootmark{e}     &  2010   &   250      & 8.1    & 2   \\
                      &         &   350      & 3.0    & 1   \\
\hline
\end{tabular}
\tablefoot{Data from:
\tablefoottext{a}{IRAS Point Source Catalog (Beichman et al. \cite{beich88}).}
\tablefoottext{b}{Akari/IRC Point Source Catalogue (Ishihara et al.
\cite{isi10}).}
\tablefoottext{c}{Akari/FIS Bright Source Catalogue (Yamamura et al.
\cite{yam10}).}
\tablefoottext{d}{This work.}
\tablefoottext{e}{Observations of Hi-GAL (Molinari et al. \cite{mol10})
retrieved from the $\textit{Herschel}$ archive.}
}
\end{table}

The corrected measurements are presented in Table~\ref{table:1}. They
enabled us to construct the infrared SED of the nebula, along with the
archived spectrum that was part of the observations carried out by the
Infrared Space Observatory (ISO) mission (Kessler et al. \cite{kessler96}).
A detailed discussion on this ISO-LWS spectrum can be found in Voors et
al. (\cite{voo00}).

The infrared SED of the nebula around AG Car obtained at different epochs
with the various instruments is shown in Fig.~\ref{agcar_sed_2dust}. All
these measurements agree very well within the uncertainties except the one
at 90 $\mu$m. This is likely due to the uncertainty on the color correction
that was stronger for the 90 $\mu$m (AKARI-FIS) data point.

A model of the dust nebula around the LBV AG Car has been carried out
in the past by Voors et al. (\cite{voo00}). They used a one-dimensional
radiative transfer code to fit both imaging and spectroscopic infrared
data. To further constrain the dust properties, we use the AKARI archive
imaging data and the new PACS and SPIRE imaging data in addition to the IRAS
imaging data and the ISO infrared spectrum. The PACS imaging allows
us to measure the nebular radius at the wavelengths of the bulk of
dust emission. To model the dust shell we only considered
the spectrum at $\lambda > $ 20 $\mu$m. The spectrum at
$\lambda < $ 12 $\mu$m comes from the central star that Voors et al.
(\cite{voo00}) fitted with a spherical non-LTE model atmosphere. For the
spectrum between $\lambda \sim$ 14 and 20 $\mu$m they argued that it
too likely comes from the central star and not from some extended source.

The two-dimensional radiative transfer code 2-Dust (Ueta and Meixner
\cite{uet03}) was used to model and interpret the dust emission spectrum
and the far-infrared images. This is a  publicly available versatile
code that can be supplied with various grain size distributions and optical
properties as well as complex axisymmetric density distributions.

It should be mentioned here that since the PACS spectral field of view
is smaller than the nebular size, the PACS spectrum was not taken into
consideration for the dust model as this spectrum is indeed fainter than
the PACS photometric points that contain the total flux of the nebula.

To model the PACS ring of dust with the code 2-Dust, it is necessary
to consider the morphology of the nebula revealed through the infrared
PACS and the optical images so as to choose the best geometric parameters
for the axisymmetric dust density distribution model. The 2-Dust code
uses a normalized density distribution function (Meixner et
al. \cite{meix02}) that is based on a layered shell model,
\begin{equation} \label{density distribution}
        \begin{split}
         \rho(R,\theta)=\left(\frac{R}{R_{\mathrm{min}}}\right)
            ^{-B\left\{1+C\sin^F\theta\left[e^{-(R/R_{\mathrm{sw}})^D}
             /e^{-(R_{\mathrm{min}}/R_{\mathrm{sw}})^D}\right]\right\}} \\
            \times\left\{1+A(1-\cos\theta)^F\left[e^{-(R/R_{\mathrm{sw}})^E}
            /e^{-(R_{\mathrm{min}}/R_{\mathrm{sw}})^E}\right]\right\}
        \end{split},
\end{equation}
where $\rho(R,\theta)$ is the dust mass density at radius \textit{R} and
latitude \textit{$\theta$}, $R_{\mathrm{min}}$ is the inner radius of the
shell, and $R_{\mathrm{sw}}$ is the superwind radius that defines the boundary
between the spherical wind and the axisymmetric superwind. The first term 
represents the radial profile of the spherical wind; the parameters A-F define
the density profile; the radial factor \textit{B} can also be a function of the
latitude through the elongation parameter \textit{C}; \textit{A}
is the equatorial enhancement parameter; the parameter \textit{F} defines
the flatness of the shell; and \textit{D} and \textit{E} are the symmetry
transition parameters that describe the abruptness of the geometrical
transition in the shell.

It should be specified that we  only consider the bright ring-nebula
for this model. No attempt has been made to  model the nebular northern extension
since it is faint in the infrared and not clearly resolved.

\begin{figure}[!]  \resizebox{\hsize}{!}{%
\includegraphics*[width=5.9cm]{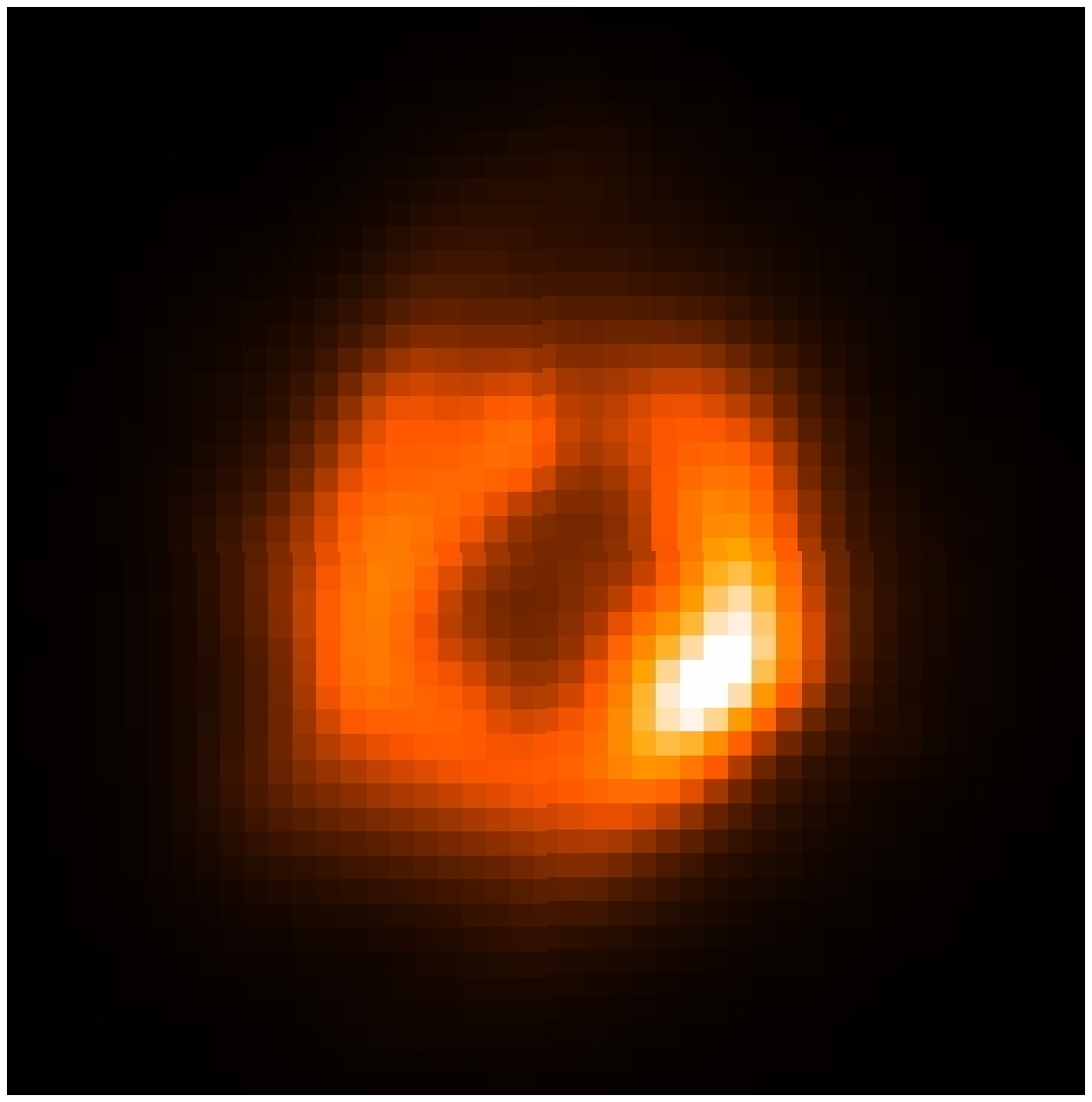}%
\includegraphics*[width=5.9cm]{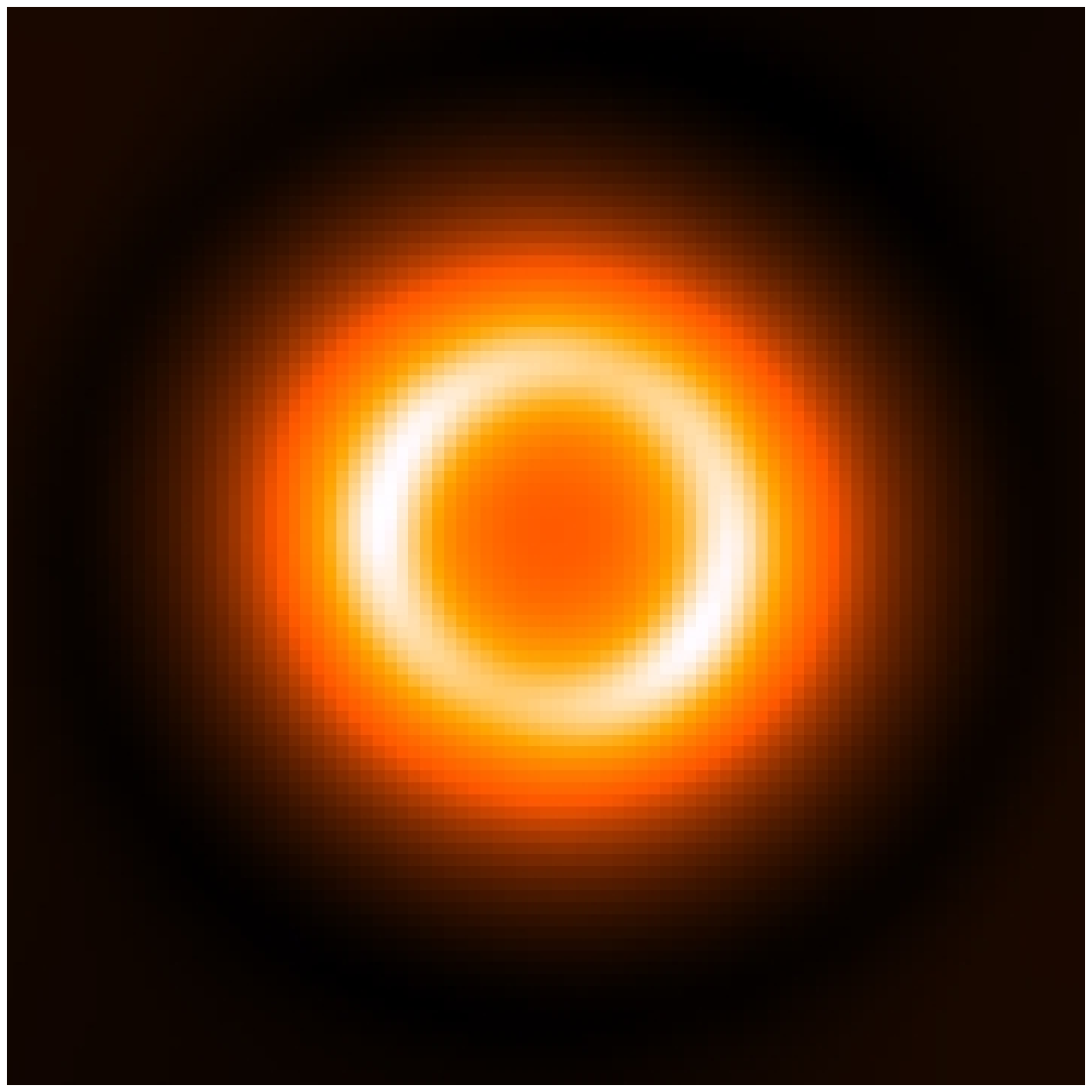}}\\%
\resizebox{\hsize}{!}{\includegraphics*{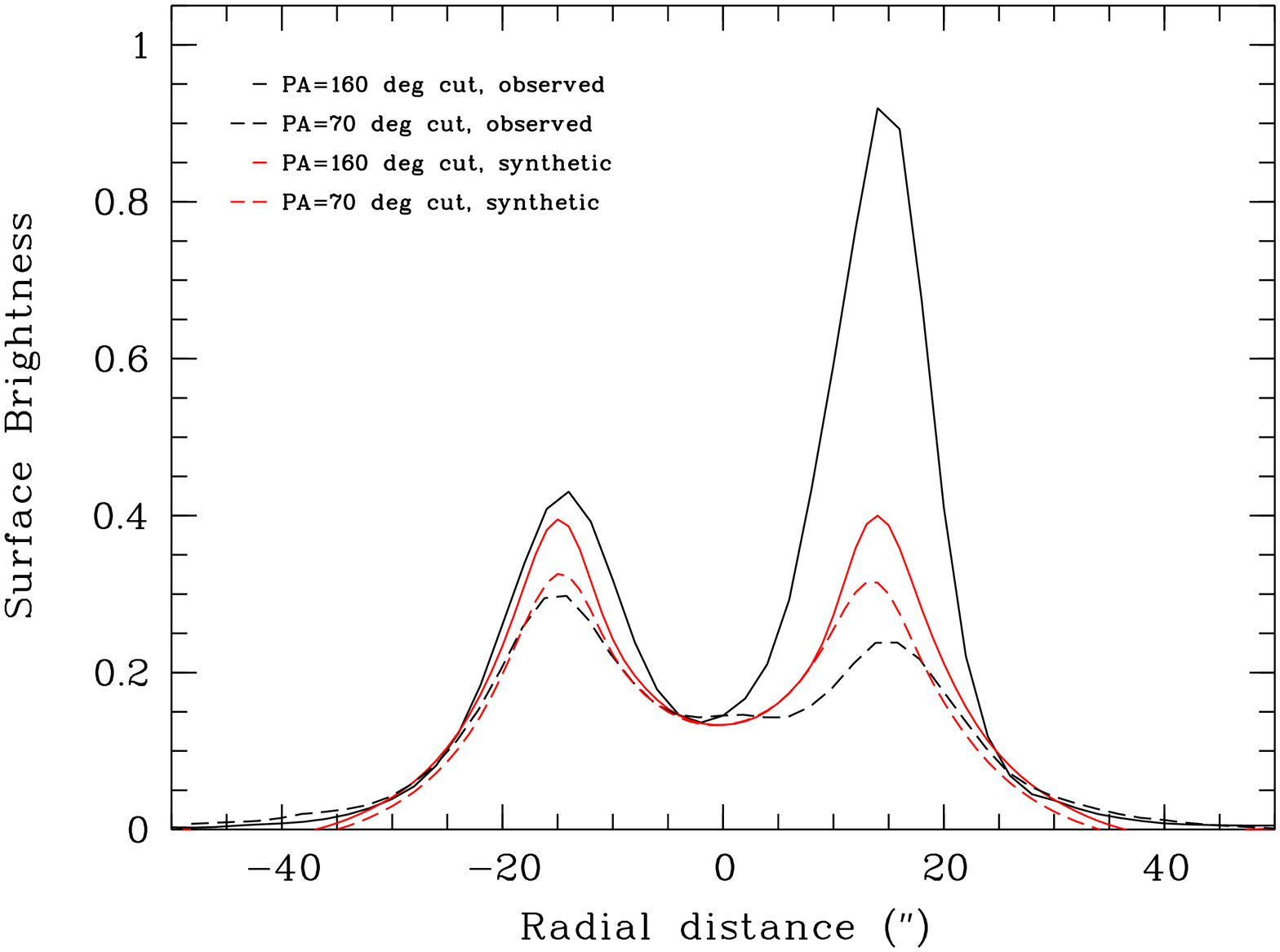}}
\caption{Top left: the $1.5\arcmin \times 1.5\arcmin$ image of the nebula
around AG Car observed with PACS at 70 $\mu$m. North is up and
east to the left. Top right: the synthetic image computed with 2-Dust
using $r_{\rm in} = 14\arcsec$ and $r_{\rm out} = 42\arcsec$ and
convolved with the PACS PSF.  Bottom: Cuts with PA=160\degr\ and 70\degr\
through the central part of the nebula at 70 $\mu$m, observed (black)
and synthetic (red) ones.}
\label{synt_cuts}
\end{figure}

A purely spherical model cannot reproduce the observed morphology of
the dust shell because  there would be too much emission at
the center of the ring with respect to the observations. A sphere with
equatorial enhancement may be considered to reproduce the difference
in intensity between the two cuts on the infrared image. To appear
nearly circular with a clear central hole in projection, a spherical
shell with equatorial enhancement must be seen at small inclination $\lesssim$
30\degr, roughly through the poles. This scheme is in agreement with the
observed global morphology. At very low inclination it is difficult to
reproduce the large difference in intensity between the two cuts in an
axisymmetric model like this one. On the other hand, by increasing the
inclination too much a strongly elliptical ring would be seen, which is not
observed. In addition to the equatorial enhancement, the structure appears
clumpy. These inhomogeneities cannot be reproduced by the present model.
The best approximate axisymmetric model for the observed ring is a sphere
with equatorial enhancement (\textit{A} $\sim$ 5) that is seen at
small inclination ($\sim$ 30\degr). The values for the other five
geometric parameters of Eq.~\ref{density distribution} are B=3,
C=0, D=0, E=0, F=1 to keep the model simple and limit the number of
free parameters.

\begin{figure}[!]
\resizebox{\hsize}{!}{\includegraphics*{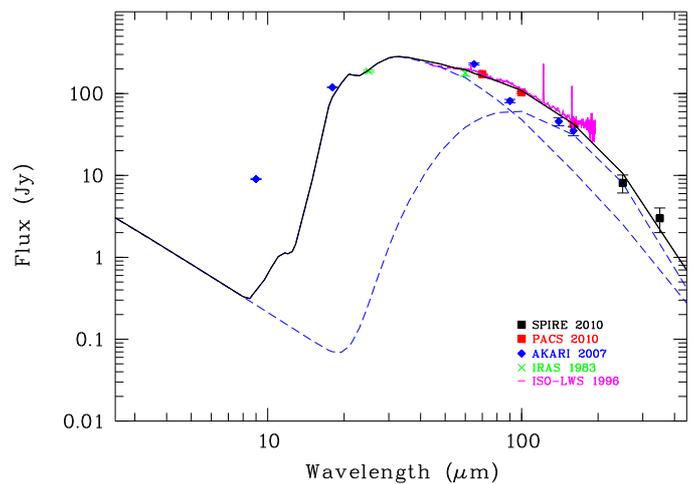}}
\caption{Infrared SED of the nebula around the LBV AG Car from data
collected at different epochs: the ISO-LWS spectrum and color-corrected
photometric measurements from IRAS, AKARI and $\textit{Herschel}$ (PACS
and SPIRE) space observatories. Results of the 2-Dust model fitting are
illustrated. Data at $\lambda < $ 20 $\mu$m are not considered in the fit.
The best fit (solid line) is achieved considering two populations of dust
grains (dashed lines).}
\label{agcar_sed_2dust}
\end{figure}

The synthetic image produced by 2-Dust and convolved with the PACS PSF is
given in the top right panel of Fig.~\ref{synt_cuts}. By comparing the PACS
image to the synthetic one, we determined the inner, $r_{\mathrm{in}} = 14\arcsec$,
and the outer, $r_{\mathrm{out}} = 42\arcsec$, radii of the dust ring. In
the bottom panel of Fig.~\ref{synt_cuts}, cuts with PA=160\degr\ and 70\degr\
through the central part of the observed and the simulated nebula are illustrated.
Although the basic morphology (radii, thickness and axisymmetry of the shell)
reproduced by the model agrees with the observed one, the intensity of the
peaks does not because no attempt was made to fit the clumps, as
mentioned earlier.

After constraining the nebular geometry we can proceed with
the model of the dust SED. For the stellar parameters we
adopted the distance D = 6 kpc, the luminosity $\log L/L_{\odot}$
= 6.1, and the temperature $T_{\mathrm{eff}} = 20000$ K (Voors et
al. \cite{voo00}, Groh et al. \cite{gro09}).
The infrared SED of AG Car (Fig.~ \ref{agcar_sed_2dust}) is too broad
to be reproduced with only one population of dust grains, so we
considered two populations of grains with the same composition but
different sizes. Voors et al. (\cite{voo00}) showed that
the dust in this nebula contains large grains, up to 40 $\mu$m in
radius, and is dominated by amorphous silicates with little
contribution from crystalline species, and more specifically pyroxenes
with a 50/50 Fe to Mg abundance. We therefore adopted a similar
dust composition. The optical constants of silicates given by Dorschner
et al. (\cite{dor95}) were used for both populations, extrapolated to a
constant refraction index in the far-ultraviolet. The size distribution
for the dust grains of Mathis et al. (\cite{mat77}, hereafter
MRN) was assumed, for each of the two populations: $n(a) \propto a^{-3.5}$
with $a_{\rm min} < a < a_{\rm max}$, where $a$ is the grain radius. The
model can be adjusted to the data by varying $a_{\rm max}$ (or $a_{\rm min}$),
which controls the 20$\mu$m / 100$\mu$m flux density ratio, and the opacity,
which controls the strength of the emission.

\begin{figure}[t]
\centering
\includegraphics[width=7cm]{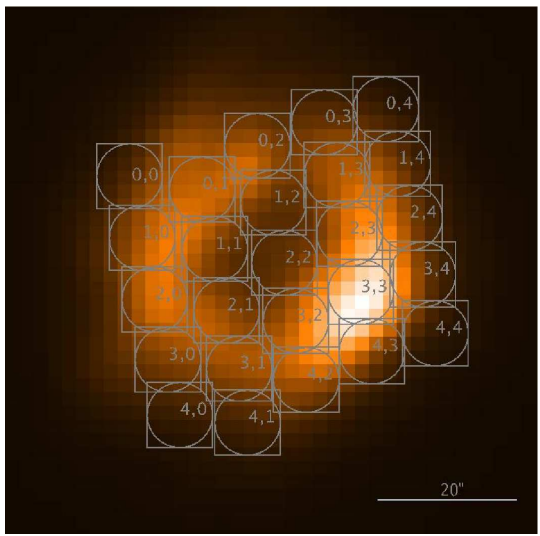}
\caption{Footprint of the PACS spectral field of view on the image 
of the AG Car nebula at 70\ \mbox{$\mu$m}. Each number pair is the label of a
specific spaxel. North is up and east is to the left.}
\label{agcar_foot}
\end{figure}

The best fit (Fig.~\ref{agcar_sed_2dust}) was achieved with the use of
the following populations of dust grains. The first is a
population of small grains with radii from 0.005 to 1 $\mu$m, which is
responsible for the emission at $\lambda < $ 40 $\mu$m. The second
is a population of large grains with radii from 1 to 50 $\mu$m, which
is responsible for the slope of the SED at $\lambda > $ 70 $\mu$m. It
should be pointed out that the large grains are necessary to reproduce
the observed infrared SED. Several attempts were made to fit the SED
using different grain sizes. They showed that $a_{\rm max}$ cannot be
very different from 50 $\mu$m for the population of large
grains. More details are given in Appendix A.

The total mass of dust derived from the modeling is $M_{\rm dust}
\sim$ 0.21 M$_{\odot}$ (0.05 M$_{\odot}$ from the small dust grains
and 0.16 M$_{\odot}$ from the large ones), with an uncertainty of
$\sim\ 20\%$. This result is in agreement with the total dust mass
found by Voors et al. (\cite{voo00}). The small grains have
temperatures from 88 K at $r_{\mathrm{in}}$ to 62 K at
$r_{\mathrm{out}}$, while the large grains have temperatures from 43 K
at $r_{\mathrm{in}}$ to 29 K at $r_{\mathrm{out}}$. Similar
results are obtained when fitting modified blackbody (BB) curves to
the SED (see Appendix B for details).

As described in the introduction, AG Car is a variable star with photometric
and spectroscopic variations. Although the data used for the dust model
are obtained at different epochs, the visual magnitude of the star is
approximately the same at the epochs of the IRAS, ISO, AKARI, and $\textit{Herschel}$
observations. In addition, AG Car varies under a roughly constant bolometric
luminosity (Sect.~\ref{sec:introduction}). By keeping the luminosity constant
and changing the radius and the temperature of the star within reasonable limits,
we do not see significant changes in the model results.

\section{Emission line spectrum}
\label{sec:emission line spectrum}

\subsection{Spectrum overview}

Figure~\ref{agcar_foot} illustrates the footprint of the PACS spectral
field of view on the image of the nebula at 70\ \mbox{$\mu$m}. This
field of view is composed of 25 (5$\times$5) spaxels, each
corresponding to a different part of the nebula, but it is not
large enough to cover the whole nebula.

\begin{figure*}
\resizebox{\hsize}{!}{\includegraphics*{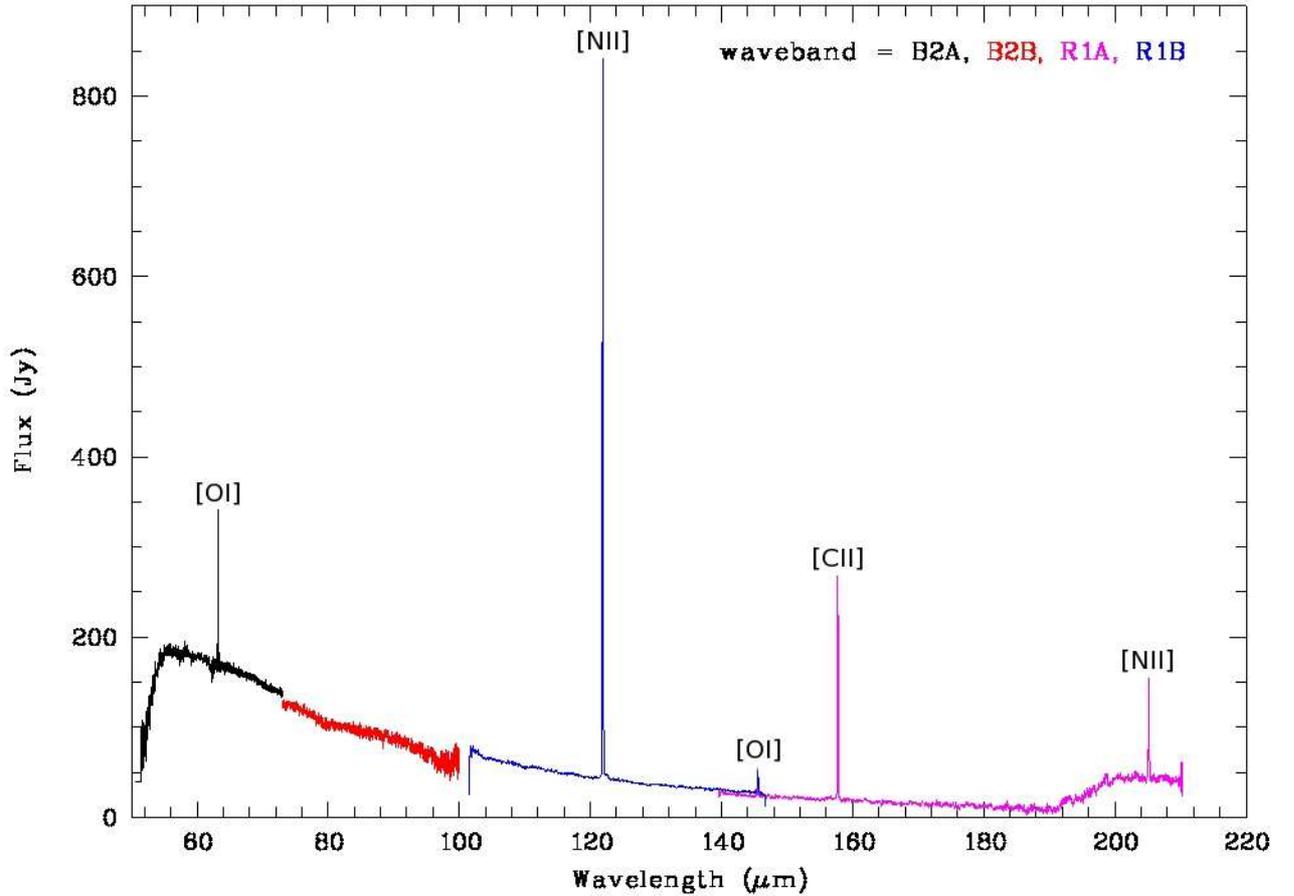}}
\caption{PACS spectrum of the nebula around AG Car, integrated over the 25
spaxels. Indicated are the lines [O{\sc i}], [N{\sc ii}], and [C{\sc ii}].
Below 55 $\mu$m the shape of the continuum results from a spectral
response function correction that has not yet been perfected. Above 190 $\mu$m
the shape results from a
light leak from the second diffraction order of the grating in the first
one. The different observing bands are indicated with different colors.
We note that the spectral resolution depends on the waveband.}
   \label{agcar_spec}
\end{figure*}

The integrated spectrum of the nebula over the 25 spaxels is
shown in Fig.~\ref{agcar_spec}. Below 55\ \mbox{$\mu$m} the shape of the
continuum results from a spectral response correction in this
range that has not yet been perfected, while above 190\ \mbox{$\mu$m}
it results from a light leak from the
second diffraction order of the grating to the first one.

The following forbidden emission spectral lines are detected
on the dust continuum: $[\ion{O}{i}]$ $\lambda\lambda$ 63, 146\ \mbox{$\mu$m},
$[\ion{N}{ii}]$ $\lambda\lambda$ 122, 205\ \mbox{$\mu$m}, and
$[\ion{C}{ii}]$ $\lambda$ 158\ \mbox{$\mu$m}. The absence of higher
ionization lines indicates that the ionization state of the gas in the
nebula around AG Car is not as high as in the case of the nebulae
around other LBVs, for example WRAY 15-751 (Vamvatira-Nakou et al.
\cite{vamv13}). This implies that the gas temperature is lower.

\subsection{Line flux measurements}

A Gaussian fit was performed on the line profiles to measure the
emission line intensities in each of the 25 spectra (Fig.~\ref
{agcar_foot}) using the Image Reduction and Analysis Facility (IRAF,
Tody \cite{tod86},\cite{tod93}). The table with these measurements is
given in Appendix C. We note that not all the lines are detected
in the outer spaxels and that all fluxes reach their highest values at
the spaxel (3,3). This spaxel corresponds to the southwestern part of
the nebula, which is the brightest part (see Sect.~\ref{sec:morphology of
the nebula}).

Maps of the line intensities were created for each of the
five detected lines in an effort to investigate whether there are
differences in the gas properties for distinct parts of the nebula.
There are only 25 spaxels, and the  coverage of the nebula is not complete,
so we cannot really see the full nebula in these ``spectroscopic images''.
The only wavelength at which we barely see the nebular ring is that of
122\ \mbox{$\mu$m}. Furthermore, maps of line intensity ratios of every
detected line to the $[\ion{N}{ii}]$ 122\ \mbox{$\mu$m} line were created.
The differences between distinct nebular regions are not significant or
convincing. Since these maps are difficult to interpret given the large
uncertainties on the fluxes and the wavelength-dependent PSF, they were
not considered in the present study.

To measure the total flux of the nebula in each line we need to use
the integrated spectrum over the 25 spaxels, but as already mentioned
above the beam size of the PACS spectrometer is smaller than the size
of the nebula. Consequently, the fluxes measured using the sum of the
25 spaxels do not correspond to the real nebular fluxes. For this
reason, the PACS spectrum was corrected using the three PACS
photometric observations. A modified BB was fitted to these data and
another one to the continuum of the PACS spectrum, for wavelengths
smaller than 190\ \mbox{$\mu$m}.  The ratio of the two curves gives
the correction factor. In other words, the spectrum was scaled to the
photometry. This factor linearly depends on the wavelength and goes
from 1.11 at 55\ \mbox{$\mu$m} to 2.12 at 185\
\mbox{$\mu$m}. This correction assumes a constant line-to-continuum
ratio and is analogous to the point source correction
applied in the pipeline to correct the effect of flux lost for a point
source.

The $[\ion{N}{ii}]$ 205\ \mbox{$\mu$m} line has a problematic
calibration in PACS. Consequently, its flux needs to be corrected
before being used in the following analysis. More precisely,
the flux in $[\ion{N}{ii}]$ 205\ \mbox{$\mu$m} is incorrect for all
PACS measurements owing to a light leak, superimposing a lot of flux
from 102.5\ \mbox{$\mu$m} at that wavelength. The relative spectral
response function (RSRF) used to reduce the data suffers from the same
light leak. Consequently, when this RSRF is applied during the
data reduction, the signal at wavelengths $\gtrsim$ 190um is divided
by a number that is too high. The continuum at these wavelengths is
irremediably lost, but provided one can ``scale-back up'' with the
right number to compensate for the exaggerated RSRF, one can recover
the line-flux.

Using instrument test data obtained on the ground with
calibration light sources set at different and known temperatures, one
can invert the problem and reconstruct the ``clean'' RSRF, i.e., in the
absence of light leak. This suffers from some defects and a large
uncertainty due to the propagation of errors and to the very low
response of the instrument at these wavelengths. A correction factor
could nevertheless be derived from it, and confirmed within a certain
margin by comparison of the line fluxes obtained for a few sources by
both PACS and SPIRE at that wavelength.  We finally found that the
measured $[\ion{N}{ii}]$ 205\ \mbox{$\mu$m} flux should be multiplied
by a correction factor of 4.2. An error of 25 \% was assumed for the
final corrected $[\ion{N}{ii}]$ 205\ \mbox{$\mu$m} flux. \footnote{This
part of the infrared spectrum of AG Car has also been observed with
SPIRE as part of the MESS program. Unfortunately, these data cannot be
used so as to have a more precise flux for the line $[\ion{N}{ii}]$
205\ \mbox{$\mu$m} because the whole ring nebula is outside of the
detector coverage owing to the geometry of the detector array and because
the observing mode was a single pointing and not a
raster map. Consequently, any attempt to recover the nebular flux
has huge uncertainty and we decided not to include the SPIRE
spectroscopic data in our study.}

The emission line measurements of the nebula integrated over
the 25 spaxels, before and after the correction for the missing flux,
are given in Table~\ref{table:2} along with the correction
factor (c.f.) at each wavelength. It should be mentioned that the flux
measurements of the three lines present in the ISO-LWS spectrum of AG
Car ($[\ion{O}{i}]$ $\lambda$ 63\ \mbox{$\mu$m}, $[\ion{N}{ii}]$
$\lambda$ 122\ \mbox{$\mu$m} and $[\ion{C}{ii}]$ $\lambda$ 158\
\mbox{$\mu$m}) do agree with the corrected values from the PACS
spectrum within the errors, showing that the correction for the
missing flux is essentially correct.

\begin{table}[h]
\caption{Line fluxes of the nebula around AG Car.} 
\label{table:2}
\centering
\begin{tabular}{l c c c c }
\hline\hline
Ion & $\lambda$  & $F\ $(25 spaxels) & c. f. & $F\ $(corrected)   \\
    &  ($\mu$m)    &($10^{-15}$ W~m$^{-2}$) &   & ($10^{-15}$ W~m$^{-2}$)  \\
\hline\hline
    $[\ion{O}{i}]$   & 63     & 7.7 & 1.17 & 9.0 $\pm$ 1.8  \\
    $[\ion{N}{ii}]$  & 122    & 23.6 & 1.64 & 38.7 $\pm$ 7.7 \\
    $[\ion{O}{i}]$   & 146    & 0.6 & 1.83 & 1.1 $\pm$ 0.3 \\
    $[\ion{C}{ii}]$  & 158    & 4.1 & 1.92 & 7.9 $\pm$ 1.6 \\
    $[\ion{N}{ii}]$  & 205    & 1.1 & 2.26\tablefootmark{a}$\times$4.2\tablefootmark{b} & 10.3 $\pm$ 2.6 \\
\hline
\end{tabular} \\
\tablefoot{
\tablefoottext{a}{Missing flux correction}
\tablefoottext{b}{PACS/SPIRE cross-calibration factor}
}
\end{table}

\subsection{Photoionization region characteristics}

The detected emission lines $[\ion{N}{ii}]$ 122, 205\ \mbox{$\mu$m}
are associated with the \ion{H}{ii} region of the nebula around AG Car.
The other three detected emission lines originate from a region of
transition between ionized and neutral hydrogen and may indicate the
presence of a photodissociation region (PDR). Their analysis is
given in the next subsection.

\subsubsection{H$\alpha$ flux}

The H$\alpha$+$[\ion{N}{ii}]$ flux from the nebula was estimated by
integrating the surface brightness over the whole nebula. Contamination
by field stars and the background was corrected and the emission from
the occulted central part extrapolated. The continuum flux from the
reflection nebula was measured in the adjacent filter, accounting for
the difference in filter transmissions. However since AG Car is a
strong emission-line star, the reflected stellar H$\alpha$ must also
be subtracted. Considering the H$\alpha$ equivalent widths measured
by Schulte-Ladbeck et al. (\cite{sch94}) and Stahl et al. (\cite{sta01})
for AG Car in 1993-1994 (i.e., accounting for $\sim$ 1.5 years of
time-delay) we estimate the final contamination due to the
reflection nebula to be 20\%. The contribution of the strong
$[\ion{N}{ii}]$ lines was then subtracted using the $[\ion{N}{ii}]$
/H$\alpha$ ratios from available spectroscopic data and the transmission
curve of the H$\alpha$+$[\ion{N}{ii}]$ filter. The conversion to absolute
flux was done with the help of the three spectrophotometric standard stars
and three planetary nebulae observed in the same filter; the conversion
factors derived from these six objects show excellent internal
agreement.

We measured $F_{0}(\mathrm{H}\alpha$) = 1.1 $\times$
10$^{-10}$ ergs~cm$^{-2}$~s$^{-1}$ uncorrected for reddening. The
uncertainty amounts to $\sim$ 20\%.  Adopting E(B$-$V) = 0.59 $\pm$ 0.03
(de Freitas Pacheco et al. \cite{pac92}), we derived
$F_{0}(\mathrm{H}\alpha$) = 4.2 $\pm$ 0.9 $\times$ 10$^{-10}$
ergs~cm$^{-2}$~s$^{-1}$ for the AG Car nebula. This flux is higher by
a factor of 2 than the fluxes measured by Stahl (\cite{sta87}), Nota et
al. (\cite{nota92}) and de Freitas Pacheco et al. (\cite{pac92}) in
1986, 1989, and 1991, but is compatible with the H$\beta$ flux measured by
Perek (\cite{per71}) in 1969 (i.e., $F_{0}(\mathrm{H}\alpha) \simeq$ 3
$\times$ 10$^{-10}$ ergs~cm$^{-2}$~s$^{-1}$ with H$\alpha$/H$\beta$ =
6 and E(B$-$V) = 0.59). The flux density from the reflection nebula is
$F_{\lambda}$ = 3.9 $\times$ 10$^{-13}$
ergs~cm$^{-2}$~s$^{-1}$~\AA$^{-1}$ at 6650~\AA\ (the central
wavelength of the continuum filter). The high value of $F_{0}(\mathrm{H}\alpha)$
we find is in agreement with the radio flux, also observed
in 1994-1995 (Duncan and White \cite{dun02}), and E(B$-$V) = 0.59

\subsubsection{Electron density}

Smith et al. (\cite{smith97_2}) found a non-constant nebular electron
density, $n_\mathrm{e}$, that varies from 600 to 1050 cm$^{-3}$ using
the optical $[\ion{S}{ii}]$ 6731/6717 ratio as an electron density
diagnostic. Their result is in agreement with those of Mitra
and Dufour (\cite{mitra90}) and Nota et al. (\cite{nota92}) who used
the same ratio as a diagnostic.

In the infrared waveband, the $[\ion{N}{ii}]$ 122/205\ \mbox{$\mu$m}
ratio is a diagnostic for the electron density of the nebula at low
density, $1\ \mathrm{cm}^{-3} \leq n_\mathrm{e}\leq
10^3\ \mathrm{cm}^{-3}$ (Rubin et al. \cite{rub94}). Considering the
values of Table~\ref{table:2}, this ratio is equal to  3.8 $\pm$ 1.2
for the whole nebula.
The package \textit{nebular} of the IRAF/STSDAS environment (Shaw \&
Dufour \cite{shaw}) was used for the calculation of the electron density.
An electron temperature, $T_\mathrm{e}$, constant throughout the nebula
and equal to 6350 $\pm$ 400 K was used for all of the following calculations.
This is the average temperature calculated by Smith et al. (\cite{smith97_2}).
The electron density is then found to be $160 \pm 90\ \mathrm{cm}^{-3}$.

The calculated electron density based on the infrared data is much
lower than the density based on the optical data. This discrepancy is
usual and has also been observed in planetary nebulae (Liu et al.
\cite{liu01}, Tsamis et al. \cite{tsam03}). When the density of a
nebula is spatially inhomogeneous, different line ratios used as density
diagnostics lead to different values of the density. This is
related to the difference in the critical density between the
lines taken into consideration for the density calculation
(Rubin \cite{rub89}, Liu et al. \cite{liu01}). The lines  
$[\ion{N}{ii}]$ 122, 205\ \mbox{$\mu$m} have
lower critical densities than the lines $[\ion{S}{ii}]$
6731, 6717 \AA,\ which means that the calculated density using the first pair of lines
is smaller than the density using the second pair (Rubin \cite{rub89}).

In the following calculations we will use our estimate of the
electron density based on infrared data because the electron
density is best determined when it is similar to the critical
density of the lines whose ratio is used as a diagnostic (Rubin et
al. \cite{rub94}). Otherwise, any attempt to calculate ionic
abundances will give incorrect results  (Rubin \cite{rub89},
Liu et al. \cite{liu01}).

\subsubsection{Ionizing flux}

To calculate the radius of the Str\"omgren sphere, $R_{S}$, and the
rate of emission of hydrogen-ionizing photons, $Q_{0}$, a steady
nonvariable star must be considered. Such an analysis can be done in
the case of a variable star like AG Car if the recombination time is
longer than the variability timescale of the ionizing star.
The recombination time is given by $\tau_{rec}= 1/n_\mathrm{e}\alpha_\mathrm{B}\ 
\mathrm{yr}$ (Draine \cite{draine11}), where $\alpha_\mathrm{B}$ is
the recombination coefficient. Using the adopted value for the electron density and
the assumed electron temperature, the recombination time was estimated
to be about 520 yr. It is much longer than the timescale of the variability
(5-10 yr) exhibited by the central star of the nebula, and this conclusion
still holds if the higher electron density derived in the optical is
considered. Therefore, an average nonvariable star is a valid approximation
in our case.

The values of $Q_{0}$ and $R_{S}$ can be determined first by using the estimated H$\alpha$
flux and second by using the radio flux density, $S_\nu$ = 268.7 mJy at 6 cm
(4.9 GHz) that was measured by Duncan and White (\cite{dun02}), adopting a
typical error of 0.5 mJy. At 4.9 GHz the nebula is optically thin and it
is assumed to be spherical with a uniform density.

The $R_{S}$ in pc is given by (Vamvatira-Nakou et al.
\cite{vamv13}) 
       \begin{equation} \label{stromgren radius}
        R_{\mathrm{S}}=3.17\left(\frac{x_e}{\epsilon}\right)^{1/3}
                      \left(\frac{n_\mathrm{e}}{100}\right)^{-2/3}
                       T_4^{(0.272+0.007\mathrm{ln}T_4)}\left
                       (\frac{Q_0}{10^{49}}\right)^{1/3},
       \end{equation}
where $Q_{0}$ (in photons $\mathrm{s}^{-1}$) using the H$\alpha$ flux is given by
       \begin{equation} \label{Q_H alpha}
        Q_{0(\mathrm{H\alpha})}=8.59\times10^{55}
                  T_4^{(0.126+0.01\mathrm{ln}T_4)}D^2F_0(\mathrm{H}_{\alpha}) \; ;
       \end{equation}
when using the radio flux it is given by
       \begin{equation} \label{Q_radio}
        Q_{0(\mathrm{radio})}=8.72\times10^{43}T_4^
                     {(-0.466-0.0208\mathrm{ln}T_4)}\left(\frac{\nu}{4.9}
                     \right)^{0.1}x_e^{-1}D^2S_{\nu} \; ,
       \end{equation}
where $x_e=n_e/n_p$ is the ratio of the electron to the proton density,
$\epsilon$ is the filling factor, $T_4=T_e/(10^4\ \mathrm{K})$, $\nu$ is
the radio frequency in GHz, $\textit{D}$ is the distance of the nebula
in kpc, $F_0(H\alpha)$ is the H$\alpha$ flux in ergs~cm$^{-2}$~s$^{-1}$,
and $S_{\nu}$ is the radio flux in mJy.

Assuming $x_e=1$ because the star is not hot enough to significantly
ionize He and $T_4=0.635$, we found that the rate of emission of
hydrogen-ionizing photons is $Q_{0(\mathrm{H\alpha})}=(1.2\ \pm\ 0.5)
\times 10^{48}~\mathrm{photons~s^{-1}}$ and
$Q_{0(\mathrm{radio})}=(1.0\ \pm\ 0.4) \times 10^{48}\
\mathrm{photons~s^{-1}}$. There is a good agreement between these two
results within the uncertainties implying that the value of E(B$-$V)
adopted for the calculations is essentially correct. The mean value
$Q_{0}=(1.1\ \pm\ 0.3)\times10^{48}\ \mathrm{photons~s^{-1}}$
corresponds to an early-B star with $T_{\mathrm{eff}}\sim 26000\
\mathrm{K}$ (Panagia \cite{pan73}), which can be considered as the
average spectral type of the star.

We also derived $R_{S}= 1.1\ \pm\ 0.4$ pc assuming $\epsilon=1$, i.e.,
that the ionized gas fills the whole volume of the nebula. The fact
that the nebula is a shell and not a sphere, with inner radius of
about $R_{\mathrm{in}}$ = 11\arcsec\ = 0.3 pc in H$\alpha$, does not
change this result because in that case the new Str\"omgren radius is
$R_{S}^{'}=(R_{S}^3+R_{\mathrm{in}}^3)^{1/3}$ = 1.1 pc. The
Str\"omgren radius is the radius of an ionization bounded nebula by
definition. In Sect.~\ref{sec:morphology of the nebula}, it was
observed that the faint part of the nebula in H$\alpha$ extends up to
0.7 pc from the central star. Moreover, the northern faint extension
discussed in that section extends up to 1 pc. The comparison of these
numbers with the estimated value of the Str\"omgren radius,
considering the uncertainties, leads to the conclusion that the
H$\alpha$ nebula may be ionization bounded. The presence of PDR
signatures in the spectrum supports this conclusion. This value
of the Str\"omgren radius is only an average value which can vary
locally depending on the density inhomogeneities. In particular,
according to the adopted morphological model, the electron density of
the shell could be higher along the equator and smaller along the
poles so that the ionizing radiation can reach the faint extensions or
bipolar lobes.

\subsubsection{Abundance ratio N/H}

Given the detected emission lines in the spectrum and the lack of
$[\ion{N}{iii}]$ 57\ \mbox{$\mu$m} and $[\ion{O}{iii}]$ 88\ \mbox{$\mu$m},
only an estimate of the N/H abundance number ratio can be made based on
the observed H$\alpha$ 6562.8 \AA, $[\ion{N}{ii}]$ 122\ \mbox{$\mu$m} and
205\ \mbox{$\mu$m} lines and considering that
      \begin{equation} \label{N/H abundance}
          \frac{\mathrm{N}}{\mathrm{H}}=\frac{\langle \mathrm{N}^{+}\rangle
             }{\langle \mathrm{H}^{+}\rangle}\ .
      \end{equation}
The flux ratios $F/F_0 (\mathrm{H}\beta)$ were calculated for the
two infrared lines of $[\ion{N}{ii}]$ with the observed values of
$\textit{F}$ from Table~\ref{table:2}. Using the dereddened H$\alpha$ flux,
a case-B recombination with $T_\mathrm{e}=6350 $ K was assumed to
calculate the H$\beta$ flux, adopting the effective recombination coefficient
equations of Draine (\cite{draine11}). The ionic abundances
$\mathrm{N}^{+}/\mathrm{H}^{+}$ were then derived using the package
\textit{nebular}. The N/H abundance number ratio was calculated to be
$(2.6\pm1.2)\times10^{-4}$, which is equivalent to a logarithmic value
of 12 + log(N/H) = 8.41 $\pm$ 0.20. Considering the errors, this value
is entirely compatible with that of Smith et al. (\cite{smith97_2}), which
is 8.27 $\pm$ 0.05. It is significantly higher than the solar value (7.83,
Grevesse et al. \cite{grev10}).

\subsubsection{Mass of the ionized gas}

An estimate of the ionized gas mass can be made from the H$\alpha$
and the radio emissions. The equations that are analytically derived
in Vamvatira-Nakou et al. (\cite{vamv13}) are used for this calculation.

For a spherical nebula the ionized mass in solar masses, taking into
account the H$\alpha$ emission, is given by
       \begin{equation} \label{ionized mass H alpha_sphere}
              M_{i(\mathrm{H\alpha})}^{\mathrm{sphere}}=57.9\frac{1+4y_{+}}{\sqrt{1+y_{+}}}T_4^{(0.471+0.015\mathrm{ln}T_4)}\epsilon^{1/2}
                   \theta^{3/2}D^{5/2}F^{1/2}_0(\mathrm{H}\alpha),
       \end{equation}
where $\theta$ is the angular radius of the nebula ($R=\theta D$)
in arcsec and $n_{\mathrm{H}^{+}}=n_{\mathrm{p}}$,
$n_{\mathrm{He}^{+}}$, and $n_{\mathrm{He}^{++}}$ are the number densities
of the ionized hydrogen, ionized helium, and doubly ionized helium,
respectively. Assuming $n_{\mathrm{He^{++}}}=0$ and denoting $y_{+}=n_{\mathrm{He^{+}}}/n_{\mathrm{H^{+}}}$, we have 
$x_{\mathrm{e}}=n_{\mathrm{e}}/n_{\mathrm{p}}\simeq1+n_{\mathrm{He^{+}}}/n_{\mathrm{H^{+}}}=1+y_{+}$
and $\mu_{+}\simeq1+4\,n_{\mathrm{He^{+}}}/n_{\mathrm{H^{+}}}=
1+4y_{+}$.

Considering now the radio flux and using the same formalism as above,
the mass of a spherical nebula in solar masses is given by 
       \begin{equation} \label{ionized mass radio_sphere}
              M_{i(\mathrm{radio})}^{\mathrm{sphere}}=5.82\times10^{-5}\frac{1+4y_{+}}{1+y_{+}}T_4^{0.175}
                    \left(\frac{\nu}{4.9}\right)^{0.05}\epsilon^{1/2}\theta^{3/2}D^{5/2}S^{1/2}_{\nu}.
       \end{equation}

In H$\alpha$ the nebula around AG Car is a shell with inner
radius $\theta_{\mathrm{in}}$ = 11\arcsec and an average outer radius
$\theta_{\mathrm{out}}$ = 20\arcsec. In the radio the nebula has approximately
the same radii (Duncan and White \cite{dun02}). In this case the mass of the
ionized shell nebula is given by
       \begin{equation} \label{ionized mass H alpha_shell}
               M_{i}^{\mathrm{shell}}=(\theta_{\mathrm{out}}^3-\theta_{\mathrm{in}}^3)^{1/2}\theta_{\mathrm{out}}^{-3/2}M_{i}^{\mathrm{sphere}} .
       \end{equation}
 The mass of the ionized shell nebula is thus
$M_{i(\mathrm{H\alpha})}^{\mathrm{shell}} =6.9\pm2.8\ \mathrm{M}_{\odot}$
and $M_{i(\mathrm{radio})}^{\mathrm{shell}}=6.4\pm2.5\ \mathrm{M}_{\odot}$,
with an average value of $M_{i}^{\mathrm{shell}} = 6.6\pm1.9\ \mathrm{M}_{\odot}$,
assuming $\epsilon=1$.  The assumption that the ionization of He is negligible
($y_{+} = 0$) was made because the central star has a temperature lower than
30000 K. This result is slightly higher, considering the uncertainties, than
the mass of $4.2\ \mathrm{M}_{\odot}$ estimated by Nota et al. (\cite{nota92},
\cite{nota95}).

\subsection{Photodissociation region characteristics}

The fine structure emission lines $[\ion{O}{i}]$ 63, 146\ \mbox{$\mu$m}
and $[\ion{C}{ii}]$ 158\ \mbox{$\mu$m} are among the most important coolants
in PDRs (Hollenbach \& Tielens \cite{holl97}). Their detection in
our spectrum may indicate the presence of a PDR in the nebula. On
the other hand, a shock, which is the result of the interaction between
the fast stellar wind and the slow expanding remnant of a previous
evolutionary phase, could also photodissociate molecules and result in
$[\ion{O}{i}]$ and $[\ion{C}{ii}]$ emission. However, the values
of the calculated ratios of $[\ion{O}{i}]$ 63\ \mbox{$\mu$m}/$[\ion{O}{i}]$
146\ \mbox{$\mu$m} and $[\ion{O}{i}]$ 63\ \mbox{$\mu$m}/$[\ion{C}{ii}]$
158\ \mbox{$\mu$m} are in agreement with the PDR models of Kaufman et al.
(\cite{kauf99}) and not with the shock models of Hollenbach and McKee
(\cite{holl89}). In particular, the ratio $[\ion{O}{i}]$ 63\ \mbox{$\mu$m}/$[\ion{C}{ii}]$
158\ \mbox{$\mu$m} is a diagnostic between PDR and shock as it is < 10
in PDRs (Tielens and Hollenbach \cite{tie85}). Consequently, based on
these ratios, we can conclude that a PDR and not a shock is present in
the nebula around the LBV AG Car and that it is responsible for the
$[\ion{O}{i}]$ and $[\ion{C}{ii}]$ emission. Photodissociation regions were detected in the
nebula that surrounds the LBV HR Car (Umana et al. \cite{uman09}) and
in the nebula around the LBV candidate HD 168625 (Umana et al.
\cite{uman10}). Later on, the infrared study of the LBV WRAY 15-751 also
revealed the presence of a PDR in the nebula that surrounds this star
(Vamvatira-Nakou et al. \cite{vamv13}).

The physical conditions in the PDR can be determined using these three
infrared lines, but because of the vicinity of the bright Carina nebula,
we have to check if these lines come entirely from the LBV nebula or if there
is a significant contribution to the measured fluxes from the background. We
therefore checked the spectra of the background taken at two different positions
on the sky and found that the lines $[\ion{O}{i}]$ 63\ \mbox{$\mu$m}, 146\ \mbox{$\mu$m}
come entirely from the nebula. However, the flux of the line $[\ion{C}{ii}]$
158\ \mbox{$\mu$m} is contaminated by the $[\ion{C}{ii}]$ foreground/background
emission. For the nebular spectrum discussed and analyzed in this section,
the background has been subtracted, as mentioned in Sect.~\ref{sec:observations
and data reduction}. Nevertheless, careful examination of the two off-source
spectra shows that the background is strong and not uniform. The difference
between the spectra of the two off positions induces an uncertainty of at
least a factor of 2 on the $[\ion{C}{ii}]$ 158\ \mbox{$\mu$m} line flux.
Hence, the measured $[\ion{C}{ii}]$ 158\ \mbox{$\mu$m} flux is unreliable
and the mass of hydrogen in the PDR based on the $[\ion{C}{ii}]$ flux
cannot be estimated.
We note that the previous conclusion about the presence of a PDR in
the nebula is still valid when background contamination is taken into account.

The structure of the PDR is described by the density of the atomic
hydrogen, $n_\mathrm{H^0}$, and the incident FUV radiation field, $G_0$,
which can be calculated using the following equation (Tielens \cite{tie05}),
where it is expressed in terms of the average interstellar radiation field
that corresponds to an unidirectional radiation field of $1.6 \times
10^{-3}$ erg cm$^{-2}$ s$^{-1}$, 
      \begin{equation}
          G_0 = 625\frac {L_{\star}\chi}{4\pi R^2}.
      \end{equation}
where $L_{\star}$ is the stellar luminosity, $\chi$ is the fraction of this
luminosity above 6 eV, which is $\sim0.7$ for an early-B star (Young
Owl et al. \cite{young02}), and $\textit{R}$ is the distance from the
star. For the PDR of the AG Car nebula, the incident FUV radiation field
is then $G_0\simeq3.7\times10^4$, considering that $L_{\star}=10^{6.1}L_{\odot}$
(Sect.~\ref{sec:dust continuum emission}) and $\textit{R}$ = 0.7 pc, which
is the radius of the ionized gas region surrounded by the PDR. This result
can be used to constrain the density of the PDR. The diagnostic
diagram of the PDR models of Kaufman et al. (\cite{kauf99}, Figs. 4 and 5)
give the ratios of the fluxes
$F_{[\ion{O}{i}]63}/F_{[\ion{C}{ii}]158}$ and
$F_{[\ion{O}{i}]145}/F_{[\ion{O}{i}]63}$
as a function of the density and the incident FUV radiation field.
By using only the latter ratio and the calculated $G_0$ and considering
the uncertainties, we can estimate the density of the PDR to be
$\log n_\mathrm{H^0}\simeq$ 3, with a large uncertainty.

To verify the consistency of the PDR analysis with the results
of the dust nebula analysis, the dust temperature, $T_\mathrm{dust}$,
can be estimated based on the radiative equilibrium, since the dust
absorbs and re-emits the FUV radiation in the far-infrared. In case of
silicates (i.e., $\beta=2$) the dust temperature is given by
(Tielens \cite{tie05})
      \begin{equation}
          T_\mathrm{dust}=50\left(\frac{1\mu\mathrm{m}}{a}\right)^{0.06}
                          \left(\frac{G_0}{10^4}\right)^{1/6} \mathrm{K\ \
                         for}\ T_\mathrm{dust} < 250\ \mathrm{K} \; .
      \end{equation}
We obtain a dust temperature of $T_\mathrm{dust}$=71 K, assuming a typical
grain size of $a = 0.1\ \mu \mathrm{m}$ because the average cross-section
is dominated by small grains. This result is in agreement with the results
of the 2-Dust model (Sect.~\ref{sec:dust continuum emission}).

\section{Discussion}
\label{sec:discussion}

The parameters of the LBV AG Car given in Table~\ref{table:3}
summarize the measurements obtained in this work along with results
taken from previous studies. The stellar parameters of luminosity,
effective temperature, and distance are from Voors et
al. (\cite{voo00}), Groh et al. (\cite{gro09}), Humphreys et
al. (\cite{hum89_2}), and this work.  The parameters for the shell
include the radii, the expansion velocity (Smith \cite{smith91}), the
kinematic age, the ionized gas electron density and the adopted
electron temperature, the abundance ratios (N/O from Smith et
al. (\cite{smith97_2}) and N/H from our study), and the measured masses
of dust and gas.

The \textit{Herschel}-PACS infrared images of the LBV AG Car reveal a
dusty shell nebula that surrounds the central star. It is a clumpy
ring with an inner radius of 0.4 pc and an outer radius of 1.2 pc. The
H$\alpha$+$[\ion{N}{ii}]$ images show a gas shell nebula that
coincides with the dust nebula, but seems to be slightly smaller and more
elliptical. The nebula has bipolar morphology, a common feature among
LBV nebulae (Weis \cite{weis01}, Lamers et al. \cite{lam01}).

The nebula around AG Car lies in an empty cavity
(Fig.~\ref{agcar_RGB}). If associated with the star, the cavity may
correspond to a previous mass-loss event when the wind of the O-type
progenitor formed a bubble, as in the case of WR stars (Marston
\cite{mar96}).  A similar case is the cavity observed around the LBV
WRAY 15-751 (Vamvatira-Nakou et al. \cite{vamv13}), though the latter
is much larger. Velocity mapping of the surrounding interstellar gas
would be needed to confirm this hypothesis and derive constraints on
the O-star evolutionary phase.

The results of our study point to a shell nebula of ionized gas and
dust, surrounded by a thin photodissociation region that is heated by
an average early-B star. The dust mass-loss rate is about (1.8 $\pm$
0.5) $\times\ 10^{-5}$ M$_{\odot}$ yr$^{-1}$, considering the duration
of the enhanced mass-loss episode that was estimated from the kinematic
age of the inner and outer radii of the nebula. Because we do not know
the total gas mass as we cannot calculate the neutral gas mass, we must
assume a gas-to-dust mass ratio in order to estimate the total
mass-loss rate. A typical value for this ratio is 100 and so the gas
mass will be $\sim$ 20 M$_{\odot}$. In the study of the nebula around
the LBV WRAY 15-751 (Vamvatira-Nakou et al. \cite{vamv13}), this ratio
was calculated to be about 40. If we assume a similar value, the gas
mass will be about 10 M$_{\odot}$, higher than but comparable to the
mass of the ionized gas. Adopting the average value, the gas mass of
the nebula around AG Car is about 15 M$_{\odot}$ with an uncertainty
of about 30\%.  The total mass-loss rate is then estimated to be (1.4
$\pm$ 0.5) $\times\ 10^{-3}$ M$_{\odot}$ yr$^{-1}$.

\begin{table}[t]
\caption{Parameters of the LBV AG Car and its shell nebula.}
\label{table:3}
\centering
\begin{tabular}{llc}\hline\hline\\[-0.10in]
Star  & log $L/L_{\odot}$                       &  6.1 $\pm$ 0.2    \\
      & $T_{\mathrm{eff}}$ (K)                  &  20000 $\pm$ 3000 \\
      & $D$ (kpc)                               &  6.0 $\pm$ 1.0    \\
Shell & \textit{r}$_{\mathrm{in}}$  (pc)        &  0.4              \\
      & \textit{r}$_{\mathrm{out}}$  (pc)       &  1.2              \\
      & ${\rm v}_{\mathrm{exp}}$ (km s$^{-1}$)  &  70               \\
      & $t_{\mathrm{kin}}$  (10$^{4}$ yr)       &  1.7              \\
      & $n_{\mathrm{e}}$ (cm$^{-3}$)            &  160 $\pm$ 90     \\
      & $T_{\mathrm{e}}$  (K)                   &  6350 $\pm$ 400   \\
      & N/O                                     &  5.7 $\pm$ 2.2    \\
      & 12+log N/H                              &  8.41 $\pm$ 0.20  \\
      & $M_{\mathrm{dust}}$  (M$_{\odot}$)      &  0.20 $\pm$ 0.05  \\
      & $M_{\mathrm{ion. gas}}$  (M$_{\odot}$)  &  6.6 $\pm$ 1.9    \\
\hline
\end{tabular}
\end{table}

It is interesting to compare this mass-loss rate that
corresponds to the period during which the ejection took place with
recent mass-loss rates.  Leitherer et al. (\cite{lei94}) found
$\dot{M}(H)$ = 0.6 $\times 10^{-5}$ to 4.0 $\times 10^{-5}$ M$_{\odot}$
yr$^{-1}$ in 1990-1992 when the star luminosity was rising, showing no
significant dependence on the luminosity phase.  Groh et
al. (\cite{gro09}) studied the fundamental parameters of AG Car during
the last two periods of minimum, 1985-1990 and 2000-2003, and
calculated a mass-loss rate from 1.5 $\times 10^{-5}$ to 6.0 $\times
10^{-5}$ M$_{\odot}$ yr$^{-1}$. The mass-loss rate during the nebula
ejection phase thus appears roughly 50 times higher than in the
present evolutionary phase.

The N/O ratio of 5.7 $\pm$ 2.2 calculated by Smith et
al. (\cite{smith97_2}) points to the presence of highly processed
material because it is much higher than the solar abundances. It is
the highest value of N/O among the known LBVs, except the case of
$\eta$ Car (Smith et al. \cite{smith98}). The 12+log N/H abundance of
8.41 $\pm$ 0.20, calculated on the basis of our observations, is
enhanced by a factor of 4.3 with respect to the solar abundance. It is lower
than the value for the LBV $\eta$ Car and higher than the values
reported for all other LBVs (Smith et al. \cite{smith98}).

\begin{figure}[t]
\resizebox{\hsize}{!}{\includegraphics*{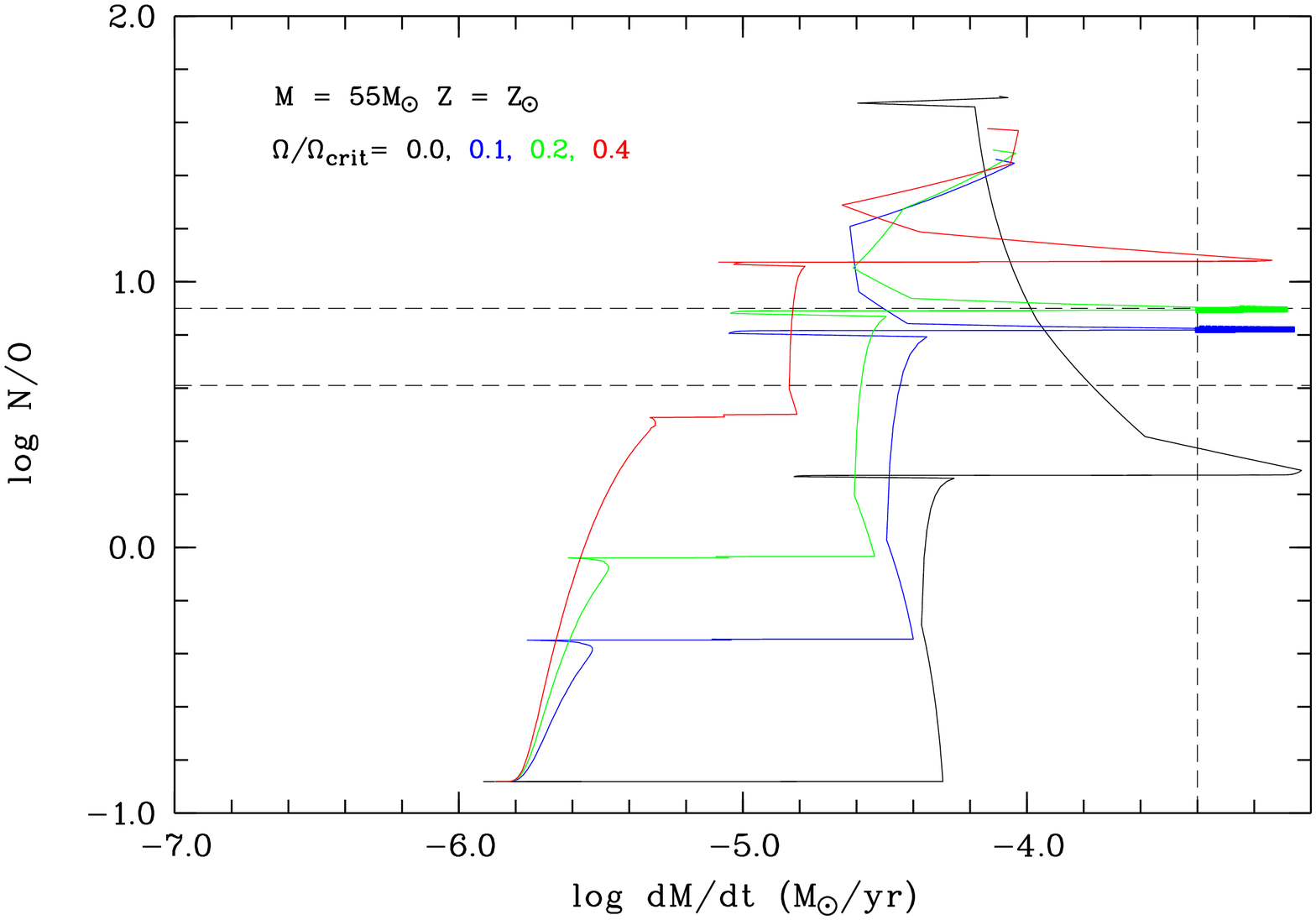}}
\caption{Evolution of the N/O surface abundance ratio as a
function of the mass-loss rate for a 55 M$_{\odot}$ star of
solar metallicity and for initial rotation rates
$\Omega/\Omega_{\rm{crit}}$ from 0 to 0.4, using the models
of Ekstr\"om et al. (\cite{eks12}). The dashed lines correspond
to the adopted value of N/O, with its errors, and the lower limit
for the mass-loss rate. The thicker lines emphasize the part of
the tracks compatible with the measurements.
For clarity, the tracks are stopped during the He burning phase
(data point n$^{\rm o}$ 195 in Ekstr\"om et al. \cite{eks12}).
}
\label{agcar_nsomdot}
\end{figure}

Groh et al. (\cite{gro09}) calculated the surface abundances
of several chemical elements at the surface of the star. The
comparison of the nebular abundances with the surface ones shows that
the N/O abundance ratio of the nebula is much lower than the surface
value of 39$^{+28}_{-18}$. As the authors mention, this is compatible
with the idea that the nebulae around massive stars contain material
that is less processed than the material of the stellar photosphere.

Smith et al. (\cite{smith97_2}), based on a detailed abundances study,
argued that the AG Car nebula was formed from material ejected during
a RSG phase.  This was also the suggestion of Voors et
al. (\cite{voo00}) based on their analysis of the dusty nebula, but
Lamers et al. (\cite{lam01}), in their study of the chemical
composition of LBVs, concluded that the ejection occurred in a blue
supergiant (BSG) phase as this can better explain the high expansion
velocity. Moreover, the problem with an ejection during a RSG phase is
the lack of luminous RSGs in the HR diagram.

Based on our observations as well as on evolutionary models
(Ekstr\"{o}m et al.  \cite{eks12}), we can constrain the evolutionary
path of the central star and the epoch at which the nebula was
ejected, using the abundance ratios, the measured mass-loss rate,
and the timescale of the ejection as constraints. The only
available abundance ratio that can be used is the N/O ratio. The N/H
abundance ratio is indeed sensitive to inhomogeneities of the nebula
(Lamers et al. \cite{lam01}). It should also be stressed that the
evolutionary models for massive stars are very uncertain at the
post-main-sequence phases as they do not include any eruptive event, which means
that the mass-loss rate recipes are poorly known (Smith
\cite{smithN_14}).

A constraint on the initial rotational velocity of AG Car can be imposed,
based on the results of Groh et al. (\cite{gro11}). In their study of
AG Car during two periods of visual minimum, they concluded that the
progenitor did not have a high initial rotational velocity, although
they measured the current projected rotational velocity to be 220 km
s$^{-1}$. Their conclusions were based on the comparison with the
evolutionary paths of Meynet and Maeder (\cite{mey03}). The
luminosity and effective temperature of the star were found to be
compatible with the evolutionary tracks of a nonrotating star with
initial mass between 40 and 60 M$_{\odot}$.

\begin{figure}[t]
\resizebox{\hsize}{!}{\includegraphics*{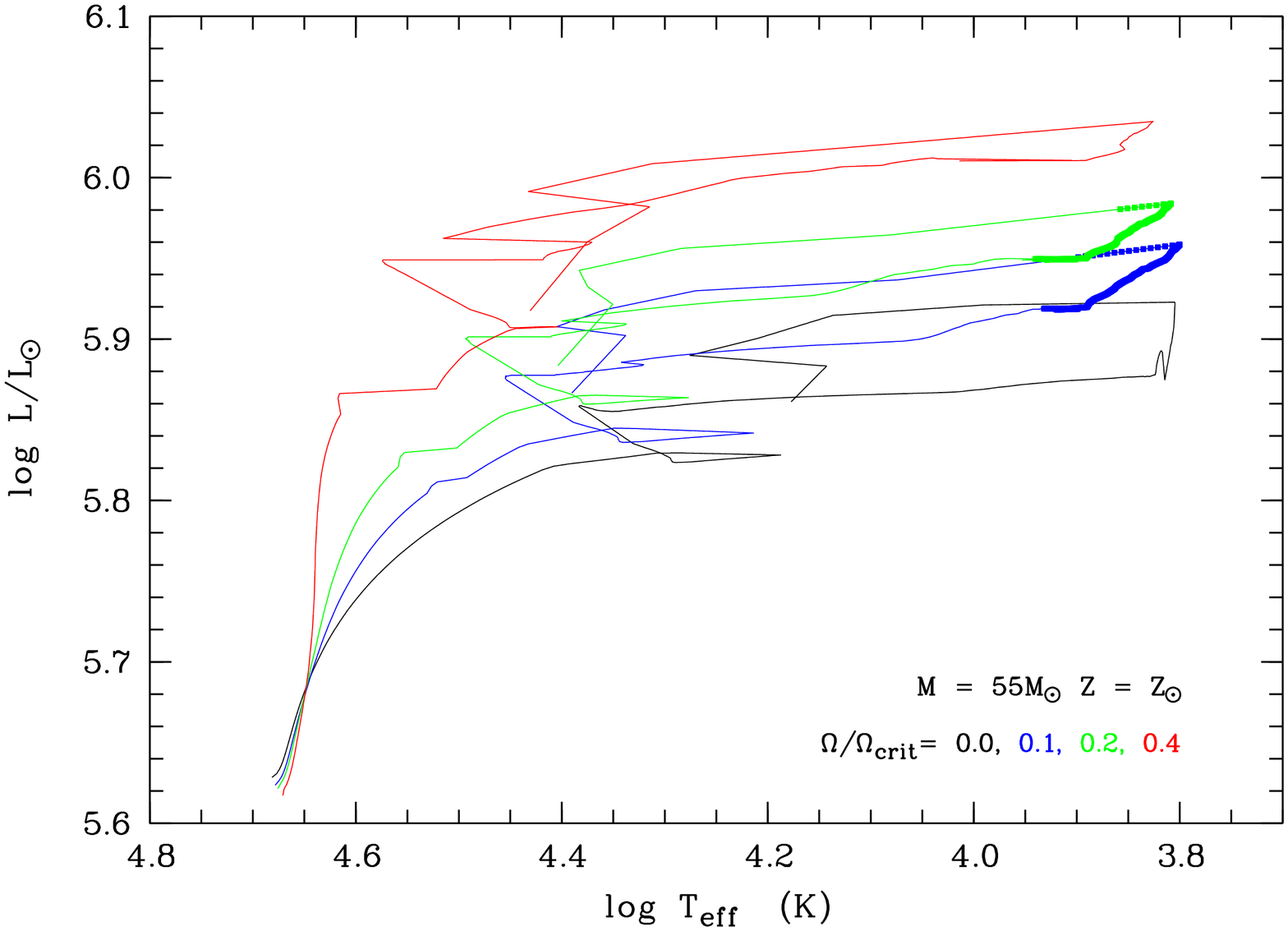}}
\caption{Evolutionary path in the HR diagram of a 55 M$_{\odot}$
star of solar metallicity and for initial rotation rates
$\Omega/\Omega_{\rm{crit}}$ from 0 to 0.4, using the models of Ekstr\"om et
al. (\cite{eks12}). The thicker lines emphasize the part of the tracks
compatible with the N/O abundance ratio and the mass-loss rate. For
clarity, the tracks are stopped during the He burning phase (data
point n$^{\rm o}$ 195 in Ekstr\"om et al. \cite{eks12}).}
\label{agcar_hr}
\end{figure}

The total mass-loss rate, estimated during the nebular ejection, is
quite high but uncertain. A lower limit of the mass-loss rate can be
considered, based on the sum of the dust mass and the ionized gas mass
that are well determined in the nebula ring. Considering the errors,
this lower limit is $\log \dot{M}$ = $-$3.4, where $\dot{M}$ is in
M$_{\odot}$~yr$^{-1}$. This result along with the nebular N/O
abundance ratio, which is assumed to be the surface abundance ratio at
the time of the ejection, were compared to the computed evolution of
the mass-loss rate versus this abundance ratio using the models of
Ekstr\"{o}m et al. (\cite{eks12}) for stars of initial masses that
correspond to the high stellar luminosity of AG Car, considering four
different cases of stellar rotation from no rotation to a rotation
rate of $\Omega/\Omega_{\rm{crit}}$=0.4. In Fig.~\ref{agcar_nsomdot},
the evolution of the mass-loss ratio versus the N/O abundance ratio is
illustrated for a 55 M$_{\odot}$ star from the models of Ekstr\"{o}m
et al. (\cite{eks12}). The measured N/O value, with its errors, and
the measured lower limit of the mass-loss rate are plotted with dashed
lines. The part of these tracks compatible with the measurements is
emphasized with thicker lines. To identify at which evolutionary phase
of the star this corresponds, the same parts of the tracks are
reported in the HR diagram (Fig.~\ref{agcar_hr}).

Our results are compatible with the evolutionary tracks of the models
of Ekstr\"{o}m et al. (\cite{eks12}) for a star of 55 M$_{\odot}$ with
solar metallicity and medium rotational velocity. In this
case, the ejection of the nebula occurs in a post-main-sequence
short-lived episode of high mass loss in agreement with the
observations. We note that in such short-lived episodes, the mass-loss
rate could be higher than computed from the model since the model
does not account for eruptive events. For a star of 57 M$_{\odot}$
the only compatible evolutionary track is the nonrotating
one. Consequently, we can conclude that the star may have a low
initial rotational velocity as suggested by Groh et al.
(\cite{gro11}). For a mass of 50 M$_{\odot}$, the only compatible
evolutionary track is the one rotating at $\Omega/\Omega_{\rm{crit}}$
= 0.4. For a mass of 60 M$_{\odot}$, the N/O ratio is reached
on the main sequence where the mass-loss rate is much smaller than our
lower limit, such that no track is compatible with both the observed N/O
ratio and a short-lived ($\lesssim$ 2 10$^4$ yr) high mass-loss
event.

A star with initial mass between 40 and 60 M$_{\odot}$ immediately
evolves to a BSG without passing through the RSG phase. It then
evolves towards the LBV and the WR phase (Meynet et
al. \cite{mey11}). Groh et al. (\cite{gro14}) performed a detailed
study on the evolutionary stages of a nonrotating star of 60
M$_{\odot}$ with solar metallicity, combining the evolutionary models
of Ekstr\"{o}m et al. (\cite{eks12}) with atmospheric models. Before
the WR phase, the evolutionary tracks of a 55 M$_{\odot}$ star
(Fig.~\ref{agcar_hr}) with little rotation are very similar to the
track of Groh et al. (\cite{gro14}) for a 60 M$_{\odot}$ without
rotation in terms of effective temperature and luminosity. Making
use of this result points to an ejection of the nebula during the LBV
evolutionary phase of AG Car and more precisely during a cool LBV
phase. Compared to the results obtained for WRAY 15-751, a lower
luminosity LBV that passed through a RSG phase where the ejection of
its nebula took place (Vamvatira-Nakou et al.  \cite{vamv13}), this
indicates that depending on their luminosity, LBV nebulae can be
ejected at different evolutionary stages. It should be mentioned that
de Freitas Pacheco et al. (\cite{pac92}) compared the AG Car nebular
properties, based on their spectroscopic observations, with the
evolutionary models available at that time, and concluded that they
were consistent with the properties of a star of 60 M$_{\odot}$ at the
beginning of the LBV phase.

The model of the dust nebula in Sect.~\ref{sec:dust continuum
emission} showed that large dust grains are necessary to reproduce the
observed infrared SED, in agreement with the results of Voors et
al. (\cite{voo00}).  This was also the case for the dust nebulae
around the LBV WRAY 15-751 (Vamvatira-Nakou et al. \cite{vamv13}) and
the yellow hypergiant Hen 3-1379, a possible pre-LBV (Hutsem\'ekers et
al. \cite{hut13}). Large grains ($a$ > 5 $\mu$m) have also been
detected in supernovae (Gall et al. \cite{gall14}).  In the case of
LBVs, the stellar temperature is most often too high for dust
formation to take place so that dust production can only happen during
large eruptions, when a pseudo-photosphere with a sufficiently low
temperature is formed. As shown by Kochanek (\cite{koch11, koch14}),
large dust grains can be produced during LBV eruptions and other
transients when the conditions of high mass-loss rate and low
pseudo-photosphere temperature are encountered. According to these
models, to produce grains larger than 10 $\mu$m, the central
star should have gone through a great outburst, with a
pseudo-photosphere temperature as low as 4000 K, i.e., much lower than
during normal eruptions. During this event, the mass-loss rate is
expected to be as high as 10$^{-2}$ M$_{\odot}$yr$^{-1}$. For AG Car,
this would require a duration of the event shorter than estimated from
the shell thickness, which is possible if the shell thickness is mostly
due to a spread in velocity (Kochanek \cite{koch11}).

\section{Conclusions}
\label{sec:conclusions}

The analysis of $\textit{Herschel}$ PACS imaging and spectroscopic
observations of the nebula around the LBV AG Car, along with optical
imaging data have been presented. The PACS images show that the dust
nebula appears as a clumpy ring. It coincides with the H$\alpha$ nebula,
but extends farther out.

The determination of the dust parameters of the nebula was performed
by dust modeling with the help of a two-dimensional radiative transfer
code.  This model points to the presence of both a small and a large
grain population of pyroxenes with a 50/50 Fe to Mg abundance. Large
grains ($a \gtrsim$ 10 $\mu$m) are needed to reproduce the
observational data.

The infrared spectrum of the nebula consists of forbidden emission
lines over a dust continuum, without the presence of any other dust
feature.  These lines reveal the presence of ionized and
photodissociation regions that are mixed with the dust. The derived
gas abundances show a strong N/O and N/H enhancement as well as a O/H
depletion, which is expected for massive evolved stars enriched with
CNO-cycle processed material.

The evolutionary path of the star and the epoch at which the nebula
was ejected were constrained using the abundances, mass-loss rate and
available evolutionary models. The results point to a nebular
ejection during a cool LBV evolutionary phase of a star with initial
mass of about 55 M$_{\odot}$ and with little rotation.

\begin{acknowledgements}
We thank the referee, Rens Waters, for his careful reading and his
constructive suggestions that greatly improved the manuscript.
C.V.N., D.H., P.R., N.L.J.C., Y.N. and M.A.T.G.
acknowledge support from the Belgian Federal Science Policy Office via
the PRODEX Programme of ESA. The Li\`ege team also acknowledges
support from the FRS-FNRS (Comm. Fran{\c c}. de Belgique). PACS has
been developed by a consortium of institutes led by MPE (Germany) and
including UVIE (Austria); KU Leuven, CSL, IMEC (Belgium); CEA, LAM
(France); MPIA (Germany); INAF-IFSI/OAA/OAP/OAT, LENS, SISSA (Italy);
IAC (Spain).  This development has been supported by the funding
agencies BMVIT (Austria), ESA-PRODEX (Belgium), CEA/CNES (France), DLR
(Germany), ASI/INAF (Italy), and CICYT/MCYT (Spain). Data presented in
this paper were analyzed using “HIPE”, a joint development by the
Herschel Science Ground Segment Consortium, consisting of ESA, the
NASA Herschel Science Center, and the HIFI, PACS and SPIRE
consortia. This research has made use of the NASA/IPAC Infrared
Science Archive, which is operated by the Jet Propulsion Laboratory,
California Institute of Technology, as well as NASA/ADS and SIMBAD
(CDS/Strasbourg) databases.
\end{acknowledgements}


\begin{appendix}

\section{Tests of the dust model}

\begin{figure}[t]
\resizebox{\hsize}{!}{\includegraphics*{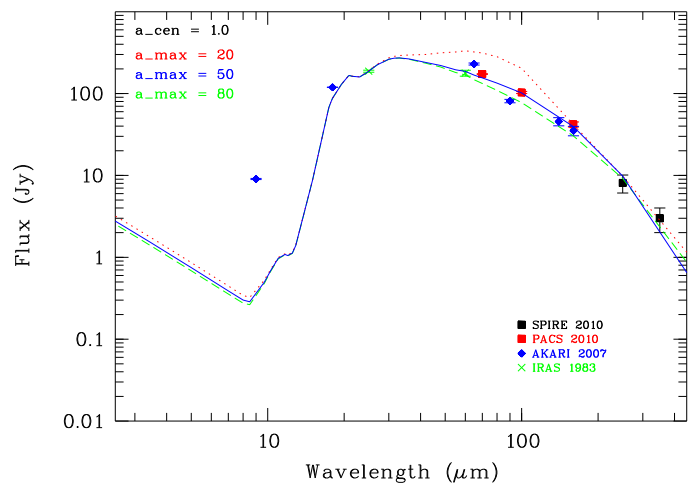}}
\caption{Fits of the far-infrared ($\lambda > 20 \mu$m) SED of
the nebula around AG Car. The dust composition (optical constants of
silicates with a 50/50 Mg-to-Fe abundance given by Dorschner et
al. \cite{dor95}) and $a_{\rm cen}$ are kept constant while the value
of $a_{\rm max}$ changes.}
\label{agcar_mod_test1}
\end{figure}

In Sect.~\ref{sec:dust continuum emission} the model of the
dust nebula is presented and the necessity of a population of large
dust grains is stressed.  Several tests have been performed using the
2-Dust code in an effort to reproduce the observed data with different
populations of dust grains from those adopted in Sect.~\ref{sec:dust
continuum emission}. Our main concern was to investigate the influence
of the dust grain size and composition on the model SED. For this
reason, we calculated many models by changing one parameter and
keeping the other ones constant. Keeping in mind  that two different populations
of grains were used to reproduce the broad observed SED in
Sect.~\ref{sec:dust continuum emission}, we consider one population of
small grains with radii $a_{\rm min} < a < a_{\rm cen}$ and one
population of large grains with radii $a_{\rm cen} < a < a_{\rm max}$.
For the dust composition, we used the optical constants given
by Dorschner et al. (\cite{dor95}) for three different abundances of
Mg to Fe, 0.5/0.5 (the model presented in Sect.~\ref{sec:dust
continuum emission}), 0.4/0.6, and 0.8/0.2. For the dust grain
sizes, the values of 20, 50, and 80 $\mu$m were considered for $a_{\rm
max}$ and the values of 0.1, 0.3, 1, and 3 $\mu$m were considered for
$a_{\rm cen}$, with the value of $a_{\rm min}$ being kept constant and
equal to 0.005 $\mu$m.

\begin{figure}[!]
\resizebox{\hsize}{!}{\includegraphics*{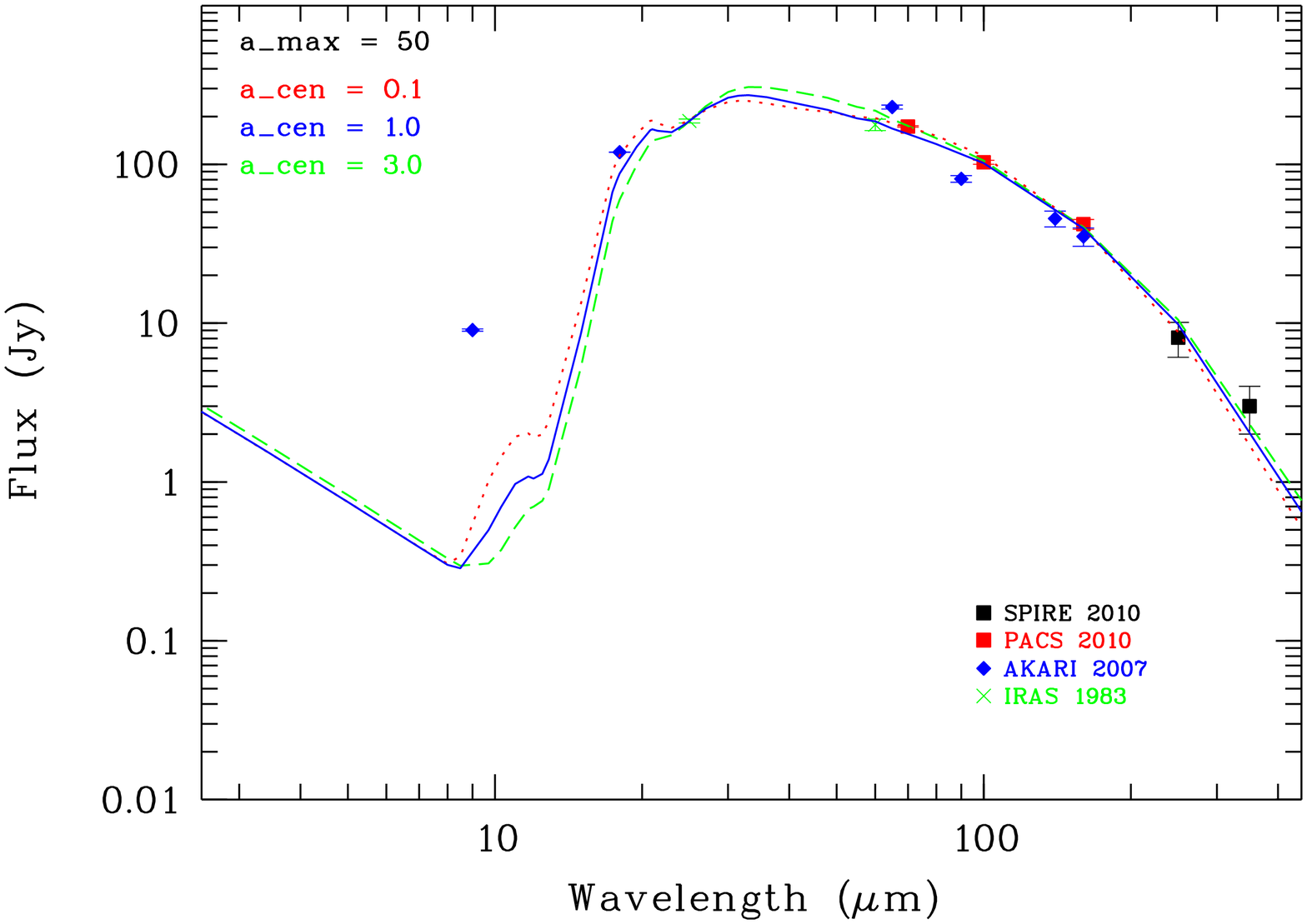}}
\caption{Fits of the far-infrared ($\lambda > 20 \mu$m) SED of
the nebula around AG Car. The dust composition (optical constants of
silicates with a 50/50 Mg-to-Fe abundance given by Dorschner et
al. \cite{dor95}) and $a_{\rm max}$ are kept constant while the value
of $a_{\rm cen}$ changes.}
\label{agcar_mod_test2}
\end{figure}

\begin{figure}[!]
\resizebox{\hsize}{!}{\includegraphics*{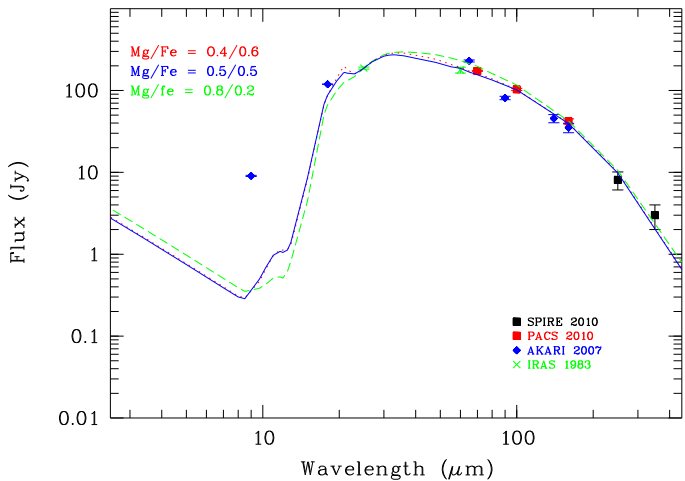}}
\caption{Fits of the far-infrared ($\lambda > 20 \mu$m) SED of
the nebula around AG Car. $a_{\rm max} = 50$ $\mu$m and $a_{\rm cen} =
1$ $\mu$m are kept constant while the dust composition changes. The
optical constants of silicates given by Dorschner et
al. (\cite{dor95}) are used for different Mg-to-Fe abundances.}
\label{agcar_mod_test3}
\end{figure}

The comparison of these tests shows that large grains are
necessary to reproduce the data. The influence of the change of
$a_{\rm cen}$ on the fit of the observed SED is almost negligible.
Furthermore, the fit depends little on the dust abundance of Mg
to Fe.  This is illustrated in Figs.~\ref{agcar_mod_test1} -
\ref{agcar_mod_test3}. In Fig.~\ref{agcar_mod_test1} three dust models
are illustrated. The dust composition is the same (optical constants
of silicates with a 50/50 Mg-to-Fe abundance given by Dorschner et
al. \cite{dor95}) and the only parameter that changes is $a_{\rm
max}$. We see that better fits to the data are achieved when large
grains are considered, in particular when $a_{\rm max}$ = 50 $\mu$m.
In all cases, when adjusting the observed flux at 250 $\mu$m, a value for $a_{\rm
max}$ that is too small ($<$ 20 $\mu$m) gives too much flux at 60-100 $\mu$m. In
Fig.~\ref{agcar_mod_test2} the comparison of three models with the
same $a_{\rm max}$ and dust composition but with different $a_{\rm
cen}$ is illustrated. The model SED depends little on $a_{\rm
cen}$. In Fig.~\ref{agcar_mod_test3} the comparison of three models
with the same grain sizes but different dust composition (in terms of
the abundance Mg to Fe) is illustrated. Again, the model SED depends
little on this dust abundance ratio, with a slightly better
adjustment for Mg/Fe = 50/50. We finally note that changing the power
law index of the grain size distribution does not significantly affect
these results.

\section{Modified blackbody fit on the SED}

\begin{figure}[!]
\resizebox{\hsize}{!}{\includegraphics*{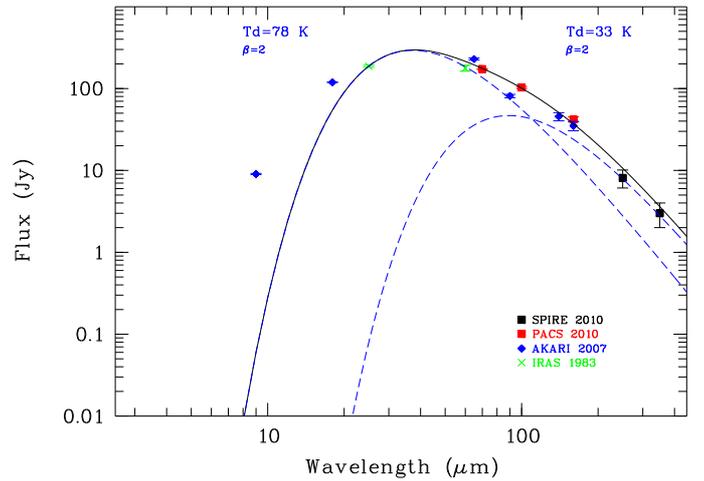}}
\caption{Same as Fig.~\ref{agcar_sed_2dust} but the infrared SED of
the nebula around the LBV AG Car is fitted by the sum of two modified
blackbody curves.}
\label{agcar_BBfit}
\end{figure}

The observed SED (Fig.~\ref{agcar_sed_2dust}) can also be reproduced
with a simpler model that is the sum of two modified BB curves
$F_{\nu}\propto B_{\nu}(T_d)\nu^{\beta}$. Only the photometric points were
considered for making the fit  illustrated in Fig.~\ref{agcar_BBfit}. The
mass of the dust can then be derived using the equation 
\begin{equation} \label{dust mass}
M_{\rm dust} = \frac{F_{\nu} \, \, D^{2}}{B_{\nu}
\left( T_{\rm d} \right) \, \kappa_{\nu}} \;,
\end{equation}
where $\kappa_{\nu}$ is the mass absorption coefficient, i.e. the absorption
cross section per unit mass, $B_{\nu}$ the Planck function and $D$ the
distance to the nebula (Hildebrand \cite{hil83}).
For this calculation, the two populations of dust grains are considered
independently. The fluxes measured at 25 $\mu$m and 250 $\mu$m are used, i.e.,
at those wavelengths where the contribution of each population dominates
(Fig.~\ref{agcar_BBfit}). At 25 $\mu$m, $\kappa_{\nu}$ = 483 cm$^{2}$g$^{-1}$
for the silicates of Dorschner et al. (\cite{dor95}). For grains of radii
smaller than the wavelengths at which dust radiates, $\kappa_{\nu}$ is roughly
independent of the radius and behaves as $\nu^{\beta}$ in the far-infrared.
However the second population of dust grains involves large grains (up to 50
$\mu$m) so that this hypothesis is no longer valid. The value of $\kappa_{\nu}$ is then
taken from the 2-Dust results where it is explicitly computed for the adopted
grain population. At 250 $\mu$m, $\kappa_{\nu}$ = 13.6 cm$^{2}$g$^{-1}$, which
is significantly higher than expected under the small-grain approximation. With
such large grains the frequency dependence of $\kappa_{\nu}$ also differs from
the $\nu^{2}$ law, being closer to $\nu^{1}$ below 100 $\mu$m. A fit with such
a composite modified BB gives $T_{\rm{dust},1}$ = 76 K and
$T_{\rm{dust},2}$ = 31 K (which is only slightly different from the values
obtained with $\nu^{2}$ only). Using $T_{\rm{dust},1}$ = 76-78 K and
$T_{\rm{dust},2}$ = 31-33 K in Eq.~\ref{dust mass}, we then derive
$M_{\rm{dust},1}$ = 0.05-0.04 M$_{\odot}$ and $M_{\rm{dust},2}$ = 0.22-0.19
M$_{\odot}$ for the small and the large grains respectively, in very good
agreement with the 2-Dust results.

\section{Emission line fluxes for each spaxel}

\begin{table*}[h]
\caption{Line fluxes in each spaxel. A dash indicates a poor S/N or a nondetection. The spatial configuration corresponds to the footprint of the PACS-spectrometer as displayed in Fig.~\ref{agcar_foot}.}
\label{table:4}
\centering
\begin{tabular}{c c| c | c | c | c | c }
\hline\hline                            \\
Ion               & $\lambda$ (band)  & $F\pm\Delta F $ & $F\pm\Delta F $ & $F\pm\Delta F $ & $F\pm\Delta F $ & $F\pm\Delta F $ \\
                  & $(\mu m)$         & ($10^{-15}$ W~m$^{-2}$)    & ($10^{-15}$ W~m$^{-2}$)    & ($10^{-15}$ W~m$^{-2}$)    & ($10^{-15}$ W~m$^{-2}$)    & ($10^{-15}$ W~m$^{-2}$)    \\
\hline\hline
                 &             & $\underline{spaxel\ 0,0}$   & $\underline{spaxel\ 0,1}$  & $\underline{spaxel\ 0,2}$  & $\underline{spaxel\ 0,3}$   & $\underline{spaxel\ 0,4}$  \\
$[\ion{O}{i}]$   & 63 (B2A)    &     -           & 0.40 $\pm$ 0.05 & 0.14 $\pm$ 0.05 & -               & -              \\
$[\ion{N}{ii}]$  & 122 (R1B)   & 0.15 $\pm$ 0.02 & 1.15 $\pm$ 0.01 & 0.85 $\pm$ 0.02 & 0.56 $\pm$ 0.02 & 0.26 $\pm$ 0.02\\
$[\ion{O}{i}]$   & 146 (R1B)   &      -          & 0.03 $\pm$ 0.01 & -               & -               & -              \\
$[\ion{O}{i}]$   & 146 (R1A)   &      -          & 0.03 $\pm$ 0.01 & -               & -               & -              \\
$[\ion{C}{ii}]$  & 158 (R1A)   &      -          & 0.21 $\pm$ 0.01 & 0.13 $\pm$ 0.01 & 0.05 $\pm$ 0.01 & -              \\
$[\ion{N}{ii}]$  & 205 (R1A)   &      -          & 0.21 $\pm$ 0.05 & 0.11 $\pm$ 0.03 & -               & -              \\
\hline
                 &             & $\underline{spaxel\ 1,0}$   & $\underline{spaxel\ 1,1}$  & $\underline{spaxel\ 1,2}$  & $\underline{spaxel\ 1,3}$  & $\underline{spaxel\ 1,4}$   \\
$[\ion{O}{i}]$   & 63 (B2A)    & 0.22 $\pm$ 0.05 & 0.65 $\pm$ 0.05 & 0.47 $\pm$ 0.05 & 0.29 $\pm$ 0.05 & -              \\
$[\ion{N}{ii}]$  & 122 (R1B)   & 0.64 $\pm$ 0.01 & 1.86 $\pm$ 0.02 & 1.49 $\pm$ 0.02 & 1.38 $\pm$ 0.01 & 0.61 $\pm$ 0.03\\
$[\ion{O}{i}]$   & 146 (R1B)   & 0.02 $\pm$ 0.01 & 0.04 $\pm$ 0.01 & 0.04 $\pm$ 0.01 & 0.02 $\pm$ 0.01 & -              \\
$[\ion{O}{i}]$   & 146 (R1A)   & 0.02 $\pm$ 0.01 & 0.04 $\pm$ 0.01 & 0.04 $\pm$ 0.01 & 0.03 $\pm$ 0.01 & -              \\
$[\ion{C}{ii}]$  & 158 (R1A)   & 0.09 $\pm$ 0.01 & 0.25 $\pm$ 0.01 & 0.24 $\pm$ 0.01 & 0.19 $\pm$ 0.01 & 0.06 $\pm$ 0.01\\
$[\ion{N}{ii}]$  & 205 (R1A)   & 0.10 $\pm$ 0.02 & 0.27 $\pm$ 0.07 & 0.29 $\pm$ 0.07 & 0.24 $\pm$ 0.06 & 0.14 $\pm$ 0.04\\
\hline
                 &             & $\underline{spaxel\ 2,0}$  & $\underline{spaxel\ 2,1}$   & $\underline{spaxel\ 2,2}$   & $\underline{spaxel\ 2,3}$  & $\underline{spaxel\ 2.4}$   \\
$[\ion{O}{i}]$   & 63 (B2A)    & 0.41 $\pm$ 0.05 & 0.26 $\pm$ 0.05 & 0.15 $\pm$ 0.05 & 0.65 $\pm$ 0.05 & 0.19 $\pm$ 0.05\\
$[\ion{N}{ii}]$  & 122 (R1B)   & 1.18 $\pm$ 0.02 & 1.43 $\pm$ 0.02 & 1.01 $\pm$ 0.02 & 1.63 $\pm$ 0.02 & 0.75 $\pm$ 0.01\\
$[\ion{O}{i}]$   & 146 (R1B)   & 0.03 $\pm$ 0.01 & 0.02 $\pm$ 0.01 & 0.02 $\pm$ 0.01 & 0.04 $\pm$ 0.01 & 0.02 $\pm$ 0.01\\
$[\ion{O}{i}]$   & 146 (R1A)   & 0.03 $\pm$ 0.01 & 0.02 $\pm$ 0.01 & 0.01 $\pm$ 0.01 & 0.05 $\pm$ 0.01 & 0.02 $\pm$ 0.01\\
$[\ion{C}{ii}]$  & 158 (R1A)   & 0.20 $\pm$ 0.01 & 0.24 $\pm$ 0.01 & 0.17 $\pm$ 0.01 & 0.30 $\pm$ 0.01 & 0.14 $\pm$ 0.01\\
$[\ion{N}{ii}]$  & 205 (R1A)   & 0.17 $\pm$ 0.04 & 0.35 $\pm$ 0.09 & 0.28 $\pm$ 0.07 & 0.33 $\pm$ 0.08 & 0.15 $\pm$ 0.04\\
\hline
                 &             & $\underline{spaxel\ 3,0}$   & $\underline{spaxel\ 3,1}$  & $\underline{spaxel\ 3,2}$   & $\underline{spaxel\ 3,3}$  & $\underline{spaxel\ 3,4}$   \\
$[\ion{O}{i}]$   & 63 (B2A)    & 0.25 $\pm$ 0.05 & 0.36 $\pm$ 0.05 & 0.60 $\pm$ 0.05 & 1.35 $\pm$ 0.06 & -              \\
$[\ion{N}{ii}]$  & 122 (R1B)   & 1.04 $\pm$ 0.02 & 1.56 $\pm$ 0.02 & 1.46 $\pm$ 0.02 & 1.68 $\pm$ 0.02 & 0.40 $\pm$ 0.01\\
$[\ion{O}{i}]$   & 146 (R1B)   & 0.03 $\pm$ 0.01 & 0.02 $\pm$ 0.01 & 0.05 $\pm$ 0.01 & 0.07 $\pm$ 0.01 & -              \\
$[\ion{O}{i}]$   & 146 (R1A)   & 0.02 $\pm$ 0.01 & 0.03 $\pm$ 0.01 & 0.04 $\pm$ 0.01 & 0.09 $\pm$ 0.01 & 0.02 $\pm$ 0.01\\
$[\ion{C}{ii}]$  & 158 (R1A)   & 0.16 $\pm$ 0.01 & 0.26 $\pm$ 0.01 & 0.32 $\pm$ 0.01 & 0.43 $\pm$ 0.01 & 0.13 $\pm$ 0.01\\
$[\ion{N}{ii}]$  & 205 (R1A)   & 0.16 $\pm$ 0.04 & 0.22 $\pm$ 0.06 & 0.23 $\pm$ 0.06 & 0.35 $\pm$ 0.09 & -              \\
\hline
                 &             & $\underline{spaxel\ 4,0}$  & $\underline{spaxel\ 4,1}$   & $\underline{spaxel\ 4,2}$   & $\underline{spaxel\ 4,3}$   & $\underline{spaxel\ 4,4}$  \\
$[\ion{O}{i}]$   & 63 (B2A)    & -               & -               & 0.58 $\pm$ 0.05 & 0.58 $\pm$ 0.06 & -              \\
$[\ion{N}{ii}]$  & 122 (R1B)   & 0.43 $\pm$ 0.02 & 0.66 $\pm$ 0.02 & 1.06 $\pm$ 0.02 & 0.49 $\pm$ 0.02 & 0.11 $\pm$ 0.02\\
$[\ion{O}{i}]$   & 146 (R1B)   & -               & -               & 0.04 $\pm$ 0.01 & 0.03 $\pm$ 0.01 & -              \\
$[\ion{O}{i}]$   & 146 (R1A)   & -               & -               & 0.04 $\pm$ 0.01 & 0.04 $\pm$ 0.01 & -\\
$[\ion{C}{ii}]$  & 158 (R1A)   & 0.04 $\pm$ 0.01 & 0.09 $\pm$ 0.01 & 0.26 $\pm$ 0.01 & 0.22 $\pm$ 0.01 & 0.04 $\pm$ 0.01              \\
$[\ion{N}{ii}]$  & 205 (R1A)   & 0.09 $\pm$ 0.02 & -               & 0.17 $\pm$ 0.04 & -               & -              \\
\hline
\end{tabular}
\end{table*}

The results of the emission line flux measurements for each spaxel
are given in Table~\ref{table:4}. The first column contains the
detected ions along with the spectral band in which the corresponding
line was measured. The following columns contain the line fluxes,
expressed in W/m$^{2}$, along with their errors. The spaxel numbers
(Fig.~\ref{agcar_foot}) are mentioned in every cell of the table. The quoted uncertainties
are the sum of the line-fitting uncertainty plus the uncertainty due
to the position of the continuum.

\end{appendix}

\end{document}